\newif\ifFULL
\newif\ifCONF

\CONFfalse
\FULLtrue

\documentclass[10pt,journal]{IEEEtran}

\usepackage[utf8]{inputenc}

\usepackage[toc]{multitoc}
\usepackage[english]{babel}
\usepackage{etex}

\usepackage{booktabs}

\usepackage{amsthm}

\usepackage{color}
\definecolor{dkgreen}{rgb}{0,0.6,0}
\definecolor{gray}{rgb}{0.5,0.5,0.5}
\definecolor{mauve}{rgb}{0.58,0,0.82}
\usepackage{listings}

\usepackage{enumerate}

\usepackage[inline,shortlabels]{enumitem}
\usepackage{caption}
\usepackage{amsfonts,amssymb,amstext,enumerate}
\usepackage[nointegrals]{wasysym}
\usepackage{array}
\usepackage[algoruled,linesnumbered]{algorithm2e}
\usepackage{graphicx}
\usepackage[table]{xcolor}
\usepackage{listing}
\usepackage{mathtools}
\usepackage{adjustbox}
\usepackage{float}
\usepackage{tikz}
\usetikzlibrary{arrows,arrows.meta,decorations.pathmorphing,decorations.footprints,fadings,calc,trees,mindmap,shadows,decorations.text,patterns,positioning,shapes,matrix,fit}
\tikzset{
    >=stealth',
    punkt/.style={
           rectangle,
           rounded corners,
           draw=black, very thick,
           minimum height=2em
           },
    pil/.style={
           ->,
           thick,
           shorten <=2pt,
           shorten >=2pt,}
}

\usepackage[
	operators,
	sets,
	adversary,
	landau,
	probability,
	notions,
	logic,
	ff,
	mm,
	advantage,
	primitives,
	events,
	complexity,
	asymptotics,
	keys,
	lambda]{cryptocode}

\usepackage{url}
\usepackage{enumitem}
\setlist{noitemsep}
\usepackage{xspace}

\usepackage{etex}
\usepackage{fontawesome}
\usepackage{wrapfig}

\usepackage{siunitx}
\graphicspath{{./figures/}}

\newtheorem{theorem}{Theorem}

\theoremstyle{definition}

\newtheorem{definition}{Definition}
\newtheorem{remark}{Remark}
\newtheorem{construction}{Construction}

\usepackage{epsdice}

\newcommand{\ranvar}[1]{\mathbf{#1}}
\newcommand{\alg}[1]{\mathsf{#1}}
\newcommand{\getsr}{{\:{\leftarrow{\hspace*{-3pt}\raisebox{.75pt}{$\scriptscriptstyle\$$}}}\:}}
\renewcommand{\set}[1]{\mathcal{#1}}

\renewcommand{\hash}{\mathsf{H}}

\renewcommand{\pk}{\mathsf{pk}}

\renewcommand{\sk}{\mathsf{sk}}
\newcommand{\btag}{\tau}
\newcommand{\genchal}{\mathsf{GenChal}}
\newcommand{\bckgen}{\mathsf{BCKGen}}
\newcommand{\snkgen}{\mathsf{SNKGen}}

\newcommand{\storagenodeNum}{k}
\newcommand{\fileproofNum}{d}
\newcommand{\storageblockNum}{m}

\newcommand{\game}{\mathbf{G}}
\newcommand{\gamePDP}{\mathbf{G}^{\mathsf{pdp}}}

\newcommand{\cIndex}{\mathcal{I}}
\newcommand{\extract}{\mathsf{E}}
\newcommand{\adversary}{\mathsf{A}}
\newcommand{\pdp}{\mathsf{PDP}}
\renewcommand{\sig}{\mathsf{SS}}
\newcommand{\chainsys}{\mathsf{BC}}

\newcommand{\file}{F}
\newcommand{\encFile}{\hat{F}}
\newcommand{\bfile}{f}
\newcommand{\encbFile}{\hat{f}}

\newcommand{\sign}{\mathsf{Sign}}

\newcommand{\createBlock}{\mathsf{Create}}
\newcommand{\selval}{\mathsf{idstr}}

\newcommand{\dealer}{\mathsf{D}}

\newcommand{\cBlockcreators}{\mathcal{BC}}
\newcommand{\blockcreator}{\mathsf{bc}}
\newcommand{\storagenode}{\mathsf{sn}}
\newcommand{\cStoragenode}{\mathcal{SN}}
\newcommand{\subFile}{\encFile}

\newcommand{\trans}{\mathsf{T}}

\newcommand{\block}{B}

\newcommand{\chal}{\mathsf{chal}}

\newcommand{\keygen}{\mathsf{KeyGen}}
\newcommand{\tagblock}{\mathsf{Tag}}
\newcommand{\genproof}{\mathsf{GenProof}}
\newcommand{\checkproof}{\mathsf{CheckProof}}

\newcommand{\setup}{\mathsf{Setup}}
\newcommand{\getFrag}{\mathsf{GetChunks}}

\newcommand{\prove}{\mathsf{Prove}}

\newcommand{\systemname}{Audita}

\newcommand{\elect}{\mathsf{Elect}}
\newcommand{\ver}{\mathsf{Ver}}

\newcommand{\Prob}[1]{\prob{#1}}
\renewcommand{\minentropy}{\mathbb{H}_{\infty}}

\newcommand{\chain}{\mathcal{C}}

\newcommand{\msg}{m}
\newcommand{\cMSG}{\mathcal{M}}
\newcommand{\Sigeufgame}{\game^{\mathsf{euf}}}

\newcommand{\rolist}{\mathcal{L}}

\newcommand{\ticketproofnum}{\ell}

\newcommand{\storetrans}{\trans_{\mathsf{store}}}
\newcommand{\jointrans}{\trans_{\mathsf{join}}}

\title{\systemname: A Blockchain-based Auditing Framework for Off-chain Storage}

\author{
\IEEEauthorblockN{
Danilo Francati\IEEEauthorrefmark{1},
Giuseppe Ateniese\IEEEauthorrefmark{1},
Abdoulaye Faye\IEEEauthorrefmark{2},
Andrea Maria Milazzo\IEEEauthorrefmark{2},\\
Angelo Massimo Perillo\IEEEauthorrefmark{1},
Luca Schiatti\IEEEauthorrefmark{2} and
Giuseppe Giordano\IEEEauthorrefmark{2}\\
}
\IEEEauthorblockA{\IEEEauthorrefmark{1}
Stevens Institute of Technology, USA\\
\{dfrancat, gatenies, aperillo\}@stevens.edu\\
}
\IEEEauthorblockA{\IEEEauthorrefmark{2}
Accenture Labs, France\\
\{abdoulaye.faye, andrea.maria.milazzo, luca.schiatti, giuseppe.giordano\}@accenture.com
}
}

\begin{document}

\maketitle

\begin{abstract}
The cloud changed the way we manage and store data. Today, cloud storage services offer clients an infrastructure that allows them a convenient source to store, replicate, and secure data online. However, with these new capabilities also come limitations, such as lack of transparency, limited decentralization, and challenges with privacy and security. And, as the need for more agile, private and secure data solutions continues to grow exponentially, rethinking the current structure of cloud storage is mission-critical for enterprises.

By leveraging and building upon blockchain’s unique attributes, including immutability, security to the data element level, distributed (no single point of failure), we have developed a solution prototype that allows data to be reliably stored while simultaneously being secured, with tamper-evident auditability, via blockchain.

The result, Audita, is a flexible solution that assures data protection and solves challenges such as scalability and privacy. Audita works via an augmented blockchain network of participants that include storage-nodes and block-creators. In addition, it provides an automatic and fair challenge system to assure that data is distributed and reliably and provably stored.


While the prototype is built on Quorum, the solution framework can be used with any blockchain platform.
The benefit is a system that is built to grow along with the data needs of enterprises, while continuing to build the network via incentives and solving for issues such as auditing and outsourcing.
\end{abstract}


\section{Introduction}\label{sec:Introduction}
A cloud storage service provides a decentralized infrastructure that allows users to store their data online.
Users pay cloud providers (such as Amazon~\cite{ADrive} or Google~\cite{GDrive}) to access the service and receive benefits such as data replication, reliability, and security.
Users' data is entirely administered by the cloud provider, which is entrusted with selecting reliable storage servers, maintaining data intact, and delivering it promptly when requested.
Also, decentralization in cloud storage is often limited, which affects data replication and results in data loss in case of natural disasters or denial of service attacks.
For example, WikiLeaks has published a highly confidential internal document~\cite{WikileaksAtlas}, showing that Amazon cloud storage (S3) has a limited number of data centers across the world.

On the other hand, the \emph{blockchain} is a technology that is fully transparent and distributed across the world. The blockchain is a sequence of public and immutable structured data (called blocks) administrated by a peer-to-peer network. It resembles the digital time-stamping system introduced by S. Haber and W. S. Stornetta \cite{Haber1990HowTT}, which was improved one year later by Bayer et al. in \cite{Bayer1993ImprovingTE}.
Today, blockchain is a building block for many technologies such as Bitcoin~\cite{nakamoto2008bitcoin} and Ethereum~\cite{EthereumWhite}.

Thanks to its properties, the blockchain is arguably an ideal infrastructure for the next generation of decentralized storage services.
However, users' data cannot be stored into blocks because the blockchain is public and immutable, and it does not scale. Encrypting or hashing relevant information and storing the results on the blockchain is also nugatory.
%
Encryption (even if information-theoretic secure) tends to ``deteriorate" over time because keys get routinely exposed or misused. Hashing data does not guarantee actual data storage. Moreover, the resultant hash provides proof of existence, which may violate privacy policies.

%

In this paper, we provide: $1$) A detailed overview of the state-of-the-art blockchain-based storage systems. $2)$ We identify the fundamental properties that such systems must satisfy, and we compare them based on these properties. $3$) We propose \systemname, the first blockchain-based decentralized storage system that satisfies all properties. It provides most of the benefits of storing data on the blockchain without actually doing it. The functionality we achieve is that {\em as long as the blockchain is growing and transactions are generated, users can be reasonably confident their data is intact even if stored off-chain}. Adding a block to the blockchain triggers an audit mechanism that implicitly verifies that a random portion of all files stored off-chain is intact. The audit is automatic and does not require file owners to be online or participate.

\subsection{Our technique}\label{par:technique}
We introduce a general technique to implement \systemname\ on top of any blockchain framework.
In our system, the blockchain is used for accountability, and data is stored off-chain in a new category of peers, called \emph{storage-nodes}, next to the standard blockchain's peers, referred to as \emph{block-creators}.
Storage-nodes provide the storage capability to save and maintain users' data, while block-creators run the protocol of the underlying blockchain.
In \systemname, a user with a file $\file$, sends a request to a server $\dealer$, called \emph{dealer}.
The file distribution phase is performed by $\dealer$ that gives, to each storage-node, a random subset of file chunks.

\begin{figure}[t!]
\centering
\includegraphics[width=\columnwidth]{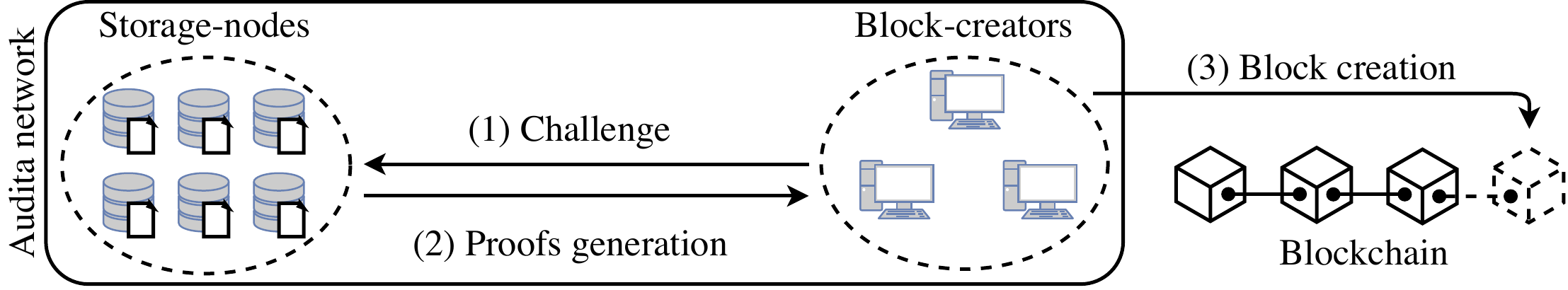}
\caption{High level protocol workflow.}
\label{fig:dealer_overview}
\end{figure}

To ensure a storage-node maintains the file subset intact (and locally stored), we leverage the blockchain and a publicly verifiable provable data possession (PDP) scheme to implement a global interactive proof system (\emph{i.e.}, prover-verifier paradigm).
A block $\block$, created by a block-creator, will be permanently added into the chain if and only if it is accompanied by a set of $\ticketproofnum$ proofs of possession $\{\pi_i\}_{i \in [\ticketproofnum]}$ (generated by $\ticketproofnum$ distinct storage-nodes over a portion of stored data).
We refer the reader to Figure~\ref{fig:dealer_overview} for a view of the protocol workflow.
In more detail, the block-creator, behaving as the verifier, challenges the storage-nodes (the provers): It contacts a random subset of storage-nodes and asks for a proof of possession $\pi_i$ (step $(1)$).
The first $\ticketproofnum$ received proofs $\{\pi_i\}_{i \in [\ticketproofnum]}$ will be used as a ``ticket'', that will give to the block-creator, the possibility to propose the candidate block $\block$ to the network (step $(2)$).
If both block $\block$ and proofs $\{\pi_i\}_{i \in [\ticketproofnum]}$ are valid, then the blockchain is extended (step $(3)$).

Thanks to its modularity, our technique allows us to implement \systemname\ on any reward-based blockchain system.
Rewards are widely used in Bitcoin and Ethereum to incentivize peers to act honestly.
\systemname\ follows the same approach. Storage-nodes and block-creators that cooperate to extend the blockchain will receive a reward in exchange for their work. The type of reward in our system can be quite arbitrary and can range from cryptocurrency-based rewards in a public blockchain to contractual agreements in a permissioned blockchain.

\subsection{Organization}
The paper is organized as follows.
Section~\ref{sec:rel-work} discusses the main blockchain-based decentralized storage systems present in the literature.
In Section~\ref{sec:prop}, we give a detailed overview of the main properties that blockchain-based decentralized storage systems must satisfy.
Section~\ref{sec:prel} introduces the notation and the required building blocks that we use throughout this paper, and Section~\ref{sec:Assumptions} defines the security assumptions we make.
In Section~\ref{sec:system-defs} we introduce \systemname\ with a detailed discussion about its properties and requirements.
Section~\ref{sec:implementation} presents our implementation and the experimental results.
Finally, Section~\ref{sec:conclusion} concludes the paper.

\section{Related work}\label{sec:rel-work}
Some recent works have focused on how to store, in a decentralized way, a publicly known file (e.g., the digital content of a library) using the blockchain.
Miller et al.~\cite{MillerJSPK14} propose Permacoin that repurposes the computational power spent in computing the Proof of Work (PoW), currently in use by many blockchain systems such as Bitcoin.
In Permacoin, the PoW involves two distinct elements to incentivize miners to store a file locally.
First, the PoW is computed over the chunks of the file.
This makes the computation an implicit form of \emph{proof of possession}.
However, this does not give any guarantee about the decentralization of the data: Miners could outsource both the file chunks and the PoW computation to a third party and jeopardize data decentralization.
For this reason, Permacoin proposes an efficient multi-use hash-based signature scheme called floating preimage signature scheme and forces miners to use their secret key during the PoW.
At a high level, miners are required to sign and hash the intermediate states of the PoW recursively.
In this way, storage outsourcing results in either exposing the secret key or decreasing the probability of generating a new valid block.
Since chunks are hashed during the computation of the PoW, Permacoin suffers from high bandwidth consumption.
Indeed, the chunks must be sent together with the PoW solution for verification.

Motivated by the need to decrease the bandwidth overhead of Permacoin, Sengupta et al.~\cite{SenguptaBRS16} propose Retricoin.
At a high level, Retricoin modifies the mining protocol of Permacoin not to involve file chunks during the computation of the PoW.
This permits to verify a candidate PoW solution without any file chunk, decreasing the network bandwidth.
Retricoin uses the PoW as a form of commitment, \emph{i.e.,} it selects the file chunk indexes that must be proven.
Then, it leverages the \emph{proof of retrievability} (PoR) scheme of Shacham and Waters~\cite{shacham2008compact} to prove the possession of that chunks.
Retricoin maintains Permacoin's objective of hindering peers from outsourcing files to third parties.
While the authors state that Retricoin is secure against outsourcing, we observe that the modification made to the mining protocol makes it susceptible to an attack.
We refer the reader to Section~\ref{sec:challenges} for more details.

Armknecht et al.~\cite{ArmknechtBKL17} propose a new PoW called EWoK (Entangled proofs of WOrk and Knowledge) that aims at  increasing the decentralization and replication of the blockchain data by including the blockchain itself into the PoW.
When EWoK is in place, the blockchain is divided into shards and the probability to compute a valid PoW solution is proportional to the number of independent shards the pool miner stores.
The mining protocol is composed of two distinct PoW phases: The first is a standard PoW whose objective is to reach consensus among the pool miners on the set of transactions to include into the block. The second one is a special PoW that is computed over the shards so that a single solution can be expected on average.
In this way, pool miners are incentivized to store different shards of the blockchain to maximize the probability of solving the second PoW.

Kopp et al. design KopperCoin~\cite{KoppBK16}, a new storage-based mining process that replaces the computationally expensive PoW.
In KopperCoin, a user can submit a special \emph{store} transaction that notifies miners of the intent to store a file in the system.
Among other things, the store transaction contains the file chunks and the public information needed to verify the proofs of retrievability computed by the miners.
Miners are free to choose which chunks to store.
The storage based mining of KopperCoin is implemented by combining a PoR scheme and a bitwise XOR-based metric $f(x,y) = x \oplus y$.
During the mining, the current block is hashed and the nearest chunk $c_j$ (measured using the bitwise XOR-based metric) is selected, \emph{i.e.,} $j = argmin_k\{H \oplus k | c_k \text{ set of chunks locally stored}\}$.
Then, a miner generates a new block $\block$ (and collects the reward) if it outputs a valid proof of retrievability for the chunk $c_j$ and $\hash(\block)\cdot2^{|j \oplus h|} \leq threshold$ where $h$ is the hash of the previous block and $threshold$ is the mining difficulty factor.
The bitwise XOR-based measure allows miners to mine proportionally to the storage offered. Indeed, the more chunks a miner stores, the higher the probability is of finding a chunk index $j$ close to $h$.
We note that KopperCoin suffers from two significant drawbacks.
First, it does not guarantee that a file is entirely stored.
This is because miners can select which chunks to store.
Hence, malicious miners could collude and choose not to store a certain portion of the file.
Second, the store transaction contains the file chunks.
This implies  that a copy of the file is stored on the blockchain.
As discussed in the introduction (Section~\ref{sec:Introduction}), this creates several  stumbling blocks (\emph{e.g.,} in relation to scalability and file deletion), and makes KopperCoin unusable in our context.

The systems proposed in~\cite{Filecoin17,Sia13,ruj2018blockstore} adopt a different approach.
They leverage the blockchain as a broadcast channel that allows users to publish orders (or contracts), rent storage and, distribute their files across the network.
Sia~\cite{Sia13} leverages a smart contract enabled blockchain to post arrangements between users and storage-nodes.
When a new arrangement is posted, the file is sent off-chain to the corresponding storage-node.
The arrangement contains several pieces of information such as: The Merkle root of the file, the expiration date, the challenge rate, and the reward for each challenge.
At specific times (determined by the challenge rate), the storage-node samples a random chunk by hashing the current blockchain block and generates a proof of possession.
The proof consists of the chunks and the list of hashes of the file's Merkle tree.
Proofs are submitted to the blockchain network and, if valid, an automatic payment is triggered that compensate the storage-node as specified in the arrangement.
A major drawback of Sia is the usage of Merkle trees.
Although they allow verification of large data, they are not meant to be used as a form of proof of possession since a verification proof must contain the challenged chunk (along with a logarithmic number of hashes).
This significantly increases the network bandwidth since chunks must be sent to other peers during the verification phase.

Protocol Labs proposes Filecoin~\cite{Filecoin17} that leverages the blockchain to implement two decentralized markets: The storage market, where users submit orders to pay storage-nodes and rent their storage, and the retrieval market, where users submit requests to retrieve a file previously stored.
Storing and retrieving files involve matching orders between users and storage-nodes.
Files are sent off-chain to storage-nodes that later generate a sequence of \emph{proofs of spacetime} to prove files were stored during the stipulated period.
Storage-nodes are paid using off-chain payment channels when they provide valid proofs.
File retrieval works similarly: Users submit retrieval requests to nodes by including payments that are  exercised off-chain only when proofs of spacetime are valid or when files have been successfully retrieved.

Blockstore~\cite{ruj2018blockstore} also employs the blockchain as a decentralized storage market, in which users can rent space from storage-nodes.
It differs from~\cite{Filecoin17,Sia13} on how the auditing tasks are handled.
Indeed, in~\cite{ruj2018blockstore}, users are responsible for their data and have to audit it by themselves or collaborate with a third party. Namely, users repeatedly challenge storage-nodes and ask for proofs of possession.

Storj~\cite{Wilkinson2016StorjAP} is an open-source project that aims to build a decentralized cloud storage service.
The core of the system is composed of a class of peers, called satellites.
They are responsible for implementing a distributed hash table (DHT) that stores all the information needed by users to upload and download files to/from the network (e.g., storage-nodes location, storage-nodes reputation, file metadata, etc.).
Additionally, they issue payments to storage-nodes, which receive an ERC20 Ethereum-based token, according to the service provided, and they are responsible for data maintenance and data replication.
As in~\cite{ruj2018blockstore}, auditing tasks are performed either by the user or a third party (in this case the satellites).
As in Sia, Storj suffers from high network bandwidth overhead since it uses Merkle trees as proofs of possession.

\section{Properties}\label{sec:prop}
In Section~\ref{sec:challenges}, we give an overview of the fundamental properties that a blockchain-based decentralized storage system must satisfy.
We make a comparison between \systemname\ and the existing systems (described in Section~\ref{sec:rel-work}) based on such properties.
In Section~\ref{subsec:sec_prop}, we report some additional challenges that must be tackled before deploying such systems in a real scenario.

\subsection{Fundamental properties}\label{sec:challenges}
We refer the reader to Table~\ref{tab:chall-table} for the comparison summary between \systemname\ and current state-of-the-art systems.

{
\newcommand{\qm}{\text{\usefont{OT1}{iwona}{m}{n}?}}
\newcolumntype{L}[1]{>{\raggedright\arraybackslash}p{#1}}
\newcolumntype{C}[1]{>{\centering\arraybackslash}p{#1}}
\newcolumntype{R}[1]{>{\raggedleft\arraybackslash}p{#1}}
\begin{table}[!t]
    \caption{
    Comparison between \systemname\ and the existing state-of-the-art systems.
    $\CIRCLE$ Property is satisfied. $\Circle$ Property is not satisfied.
    $\LEFTcircle$ Property is partially satisfied.
    Question mark $\qm$ indicates insufficient implementation details.
    The highlighted systems present the drawback of either store file chunks on the blockchain or send them across the network for verification purposes.
    }
    \label{tab:chall-table}
\begin{center}
\begin{tabular}{ R{2cm} C{0.95cm} C{1.35cm} C{0.95cm} C{1.4cm}  }
  \toprule
  & PoW-free & Decentralized auditing & Fair auditing & Resilient to outsourcing \\
  \midrule
  \rowcolor{lightgray}
  Permacoin~\cite{MillerJSPK14} & $\Circle$ & $\CIRCLE$ & $\Circle$ & $\CIRCLE$ \\
  Retricoin~\cite{SenguptaBRS16} & $\Circle$ & $\CIRCLE$ & $\Circle$ & $\Circle$\\
  EWoK~\cite{ArmknechtBKL17} & $\Circle$ & $\CIRCLE$ & $\Circle$ & $\CIRCLE$\\
  \rowcolor{lightgray}
  KopperCoin~\cite{KoppBK16} & $\CIRCLE$ & $\CIRCLE$ & $\Circle$ & $\Circle$ \\
  \rowcolor{lightgray}
  Sia~\cite{Sia13} & $\CIRCLE$ & $\CIRCLE$ & $\LEFTcircle$ & $\Circle$ \\
  Filecoin~\cite{Filecoin17} & $\CIRCLE$ & $\CIRCLE$ & $\LEFTcircle$ & $\qm$ \\
  BlockStore~\cite{ruj2018blockstore} & $\CIRCLE$ & $\Circle$ & $\LEFTcircle$ & $\Circle$ \\
  \rowcolor{lightgray}
  Storj~\cite{Wilkinson2016StorjAP} & $\CIRCLE$ & $\Circle$ & $\LEFTcircle$ & $\Circle$\\
  \midrule
  \systemname\ (Sec.~\ref{sec:implementation}) & $\CIRCLE$ & $\CIRCLE$ & $\CIRCLE$ & $\CIRCLE$ \\
  \bottomrule
 \end{tabular}
\end{center}
\end{table}
}

\paragraph*{PoW-free}\label{subsec:storagerequirement}
The current decentralized storage systems can be divided into two distinct categories according to how they prove the integrity of the files: PoW-based and PoW-free.
Systems such as~\cite{MillerJSPK14,SenguptaBRS16,ArmknechtBKL17} are PoW-based: Nodes are required to spend a significant amount of their computational power to compute the PoW.\footnote{Retricoin~\cite{SenguptaBRS16} leverages PoW to select the chunks indexes to challenge.}
On the other hand,~\cite{KoppBK16,Sia13,Filecoin17,ruj2018blockstore,Wilkinson2016StorjAP} do not rely on PoW at all: Nodes are only required to generate proofs of possession.
A decentralized storage system should not make use of PoW for two main reasons: It results in a significant waste of electricity, and it excludes from the network nodes with limited computational power but available storage space.
\systemname\ belongs to the PoW-free category: A storage-node must only provide its storage and have access to modest network connectivity to interact with other peers. Note, however, that block-creators could still deploy PoW as their consensus mechanism if deemed suitable.

\paragraph*{Decentralized auditing and fairness}
Data auditing is the core of a decentralized storage system.
Once users upload their files, they expect to receive assurance that files are stored correctly, and their integrity is preserved.
The auditing process is composed of two distinct aspects that we call \emph{decentralization} and \emph{fairness}.

Decentralization refers to how the system audits (and distributes) the data.
Resorting to the file owners for auditing activities (via PDP or PoR) is not acceptable in a decentralized scenario. Users can be offline or fail to trigger audit events.  Auditing must be decentralized and automatically performed by the system.
Blockstore~\cite{ruj2018blockstore} and Storj~\cite{Wilkinson2016StorjAP} are centralized: The user authorizes a third party to audit its data by continually challenging storage-nodes.
On the other hand,~\cite{MillerJSPK14,SenguptaBRS16,ArmknechtBKL17,KoppBK16,Sia13,Filecoin17} are decentralized.
In particular,~\cite{MillerJSPK14,SenguptaBRS16,ArmknechtBKL17} are PoW-based, and nodes challenge themselves by computing the PoW.
A similar consideration holds for KopperCoin~\cite{KoppBK16} where, instead of the PoW, nodes execute a storage-based mining algorithm.
Lastly, in~\cite{Sia13,Filecoin17}, storage-nodes regularly generate proofs of possession to fulfill the agreement established with the user.

Fairness refers to the ability to uniformly audit every part of the file uploaded.
In other words, the auditing is fair when each file chunk has the same probability of being checked.
Systems such as~\cite{MillerJSPK14,SenguptaBRS16,ArmknechtBKL17,KoppBK16} are {\em not} fair under this definition.
In KopperCoin~\cite{KoppBK16}, storage-nodes are free to select which chunks to store and could merely choose not to preserve some of them.
While, in~\cite{MillerJSPK14,SenguptaBRS16,ArmknechtBKL17}, the probability of challenging a node is proportional to its hash power since PoW is used to prove storage integrity.

Instead, systems~\cite{Sia13,Filecoin17,ruj2018blockstore,Wilkinson2016StorjAP} partially satisfy fairness because file owners are in charge of selecting storage nodes and challenge rates in advance.

\systemname\ provides decentralized auditing that is decentralized and fair.
Storage-nodes and their respective integrity challenges are uniformly sampled by hashing the output of the election phase that determines the block-creator.

\paragraph*{Resilience to outsourcing}
Outsourcing refers to the possibility for a node to store data within a cloud storage service (\emph{e.g.}, Amazon, Google, Microsoft, etc.) to release its memory.  If several nodes outsource their data, decentralization will not be guaranteed.
A reliable decentralized storage service should tackle this issue by hindering peers that outsource their data.
As an example, systems such as~\cite{MillerJSPK14,ArmknechtBKL17} force peers to use their local storage by including both the data and the peer's secret key in the PoW process.
On the other hand,~\cite{Sia13,Wilkinson2016StorjAP,KoppBK16,ruj2018blockstore} do not propose any defense against outsourcing.
Filecoin~\cite{Filecoin17} will block outsourcing by leveraging a
\emph{proof of spacetime} and a \emph{time-bounded proof of replication}, but details about these primitives are still work in progress.
Lastly, Sengupta et al.~\cite{SenguptaBRS16} state that Retricoin is secure against outsourcing.
However, we note that file chunks are not directly hashed during the computation of the PoW (as it is done in Permacoin). Instead, the PoW involves only the indexes of the chunks that are  challenged.
Thus, a malicious miner could simply compute the PoW independently, and have the proof of retrievability generated by an external server.


\systemname\ proposes a new solution for the storage outsourcing problem by making it financially inconvenient. 
A block-creator will eventually ask and wait from storage-nodes $\ticketproofnum$ proofs of possession $\{\pi_i\}_{i \in \ticketproofnum}$.
A reward is only assigned to the first $\ticketproofnum$ storage-nodes that provide such proofs.
This consistently puts storage-nodes in a highly competitive state.
The reason is that block-creators have the incentive to include in the winning ticket the first $\ticketproofnum$ proofs $\pi_i$ received to avoid any delay in getting the reward.
If a storage-node decides to outsource its data, it will need to contact an external server to retrieve $\pi_i$.
The assumption we make here is that this will cause network delays which decreases the probability of getting the reward.
The advantage of our technique is that it does not involve any intensive computational task from the storage-nodes' side.
Note that a similar approach is used by Bowers et al.~\cite{bowers2011tell}. They show that the response time permits to verify that a server is storing a file in a fault-tolerant manner.

\subsection{Additional challenges}
\label{subsec:sec_prop}
In this section we discuss about problems that must be tackled before deploying a blockchain-based decentralized storage system.

\paragraph*{Incentivization}
In a decentralized storage service, peers offer their local storage space to build the network storage infrastructure.
Hence, one critical aspect is how to incentivize peers to join the network.
Peers must be persuaded to provide their storage and keep data intact.
Every existing system ~\cite{MillerJSPK14,SenguptaBRS16,ArmknechtBKL17,Sia13,Wilkinson2016StorjAP,Filecoin17,KoppBK16,ruj2018blockstore} uses rewards to encourage peers to participate in the storage process honestly.
In more detail, in~\cite{MillerJSPK14,SenguptaBRS16,ArmknechtBKL17,KoppBK16}, the system automatically generates rewards to peers that correctly store the file.
As an example,~\cite{MillerJSPK14,SenguptaBRS16,ArmknechtBKL17} are conceived to work with Bitcoin, and peers receive a bitcoin coinbase transaction as their reward.
A different approach is adopted by~\cite{Sia13,Filecoin17,ruj2018blockstore,Wilkinson2016StorjAP}: Users pay nodes in exchange for their storage space.
In \systemname, at each timestamp, a block-creator asks storage-nodes for proofs of possession on their respective file chunks. A reward is assigned only to the first $\ticketproofnum$ storage-nodes that answer correctly.
The goal of \systemname\ is to incentivize storage-nodes (1) to store file chunks faithfully {\it and} (2) to actively provide valid proofs of possession. Indeed, unlike other systems, \systemname\ creates a strong incentive for storage-nodes to compete with each other and provide valid cryptographic proofs promptly.

\paragraph*{Malicious users}\label{subsec:Untrusteduser}
The incentive that a blockchain-based storage system provides to invite nodes to participate can become a double-edged sword.
Clients and nodes can collude to take advantage of the network.
As mentioned in~\cite{Filecoin17}, a malicious user can generate a large file through a program.
By sharing this program with a node, the latter can free its storage space but still claim to be storing the file.
Specifically, let us assume that a malicious user requests to store a file $\file$ generated in the following way:
\begin{itemize}
\item Choose a PRF $G$ with key $s \getsr \bin^*$.
\item Create an arbitrary long file $\file = \{\bfile_1, \ldots, \bfile_n\}$ where $\bfile_i = G_s(i)$.
\item Make a request to store the file $\file$ in the network.
\item Collude by sharing the $G_s$ with the nodes that are storing the file.
\end{itemize}
If the data owner spends less money to store his file than the rewards the storage-node gets, then their collusion will generate significant earnings.
A system that could suffer from this problem is KopperCoin~\cite{KoppBK16}.
In KopperCoin a user, that wants to store a file for a fixed period must destroy a fixed amount of coins.
The longer the period, the higher the number of coins to destroy.
On the other side, storage-nodes will earn some new coins that are included in a coinbase transaction.
The authors do not mention any relationship between the number of coins to destroy and the amount contained into the coinbase transaction.
Hence, KopperCoin could be susceptible to the above attack if the amount paid by the user is lower than the expected rewards that a storage-node can earn.
On the other hand,~\cite{MillerJSPK14,SenguptaBRS16,ArmknechtBKL17,Sia13,Wilkinson2016StorjAP,ruj2018blockstore} are not susceptible to this attack.
Miller et al.~\cite{MillerJSPK14}, and Sengupta et al.~\cite{SenguptaBRS16} assumes that a {\it trusted} dealer owns the file $\file$.
Hence, the attack is out of scope since $\file$ comes from a trusted party.
A similar argument holds for~\cite{ArmknechtBKL17}: $\file$ is the Bitcoin blockchain which is publicly verifiable, and thus implicitly trusted.
Filecoin~\cite{Filecoin17} mentions that their \emph{time-bounded proof of replication} and \emph{proof of spacetime} prevent the attack, but their solution is still being worked out.
Lastly, in~\cite{Sia13,Wilkinson2016StorjAP,ruj2018blockstore}, the amount paid by the user is precisely the amount that the storage-nodes will earn during the storage process.
We stress that this approach applies to most of the systems, including \systemname.


\paragraph*{Recovery}\label{subsec:retrievability}
A decentralized storage system must allow users to retrieve the file from the network.
Unfortunately, merely deploying PDP and PoR is not sufficient.
These primitives were designed to challenge cloud storage servers, but cannot help if providers misbehave or simply disappear \cite{AccountableStorage}.
Works such as~\cite{SenguptaBRS16,ArmknechtBKL17,Sia13,ruj2018blockstore} do not provide any insight on how the file can be retrieved.
Permacoin~\cite{MillerJSPK14} assumes that a portion of altruistic peers may return the file to the user.
However, without any incentives, the number of altruistic peers could slowly decrease over time, making the file unretrievable.
On the other hand,~\cite{Wilkinson2016StorjAP,Filecoin17} propose two similar solutions that rely on the same idea: Clients pay storage-nodes to retrieve their files as an incentive.
However, they do not describe how to fairly exchange the payment with the content of the file. Thus, in principle, the storage-node could receive the payment and disappear from the network.
The solution of KopperCoin~\cite{KoppBK16} involves a $2$-out-of-$2$ multisignature.
The user and the storage-node generates a transaction which includes a payment for the storage-node and two collaterals, one for each of the parties. Honest parties get their collateral back; in particular, when the user receives the file requested, then the multisignature is used to pay the storage-node and return the two collaterals to the respective parties.

\systemname~  relies on similar techniques and leverages smart contract enabled fair exchange protocols~\cite{dziembowski2018fairswap} to implement a reliable file recovery mechanism.

\section{Preliminaries}\label{sec:prel}

\subsection{Notation}\label{subsec:notation}
We use the notation $[n] = \{1, \ldots, n\}$.
Capital boldface letters (such as $\ranvar{X}$) are used to denote random variables, small letters (such as $x$) to denote concrete values, calligraphic letters (such as $\set{X}$) to denote sets and serif letters (such as $\alg{A}$) to denote algorithms.
For a string $x \in \bin^*$, we use $|x|$ to denote its length; if $\set{X}$ is a set, $|\set{X}|$ is the number of elements in $\set{X}$.
If $\alg{A}$ is an algorithm, we use $y = \alg{A}(x)$ to denote the run of $\alg{A}$ on input $x$ and output $y$; If $\alg{A}$ is a randomized algorithm we write $\alg{A}(x;r)$ to denote the run of $\alg{A}$ on input $x$ and uniform randomness $r$).
We sometimes write $y \getsr \alg{A}(x)$ to denote the output $y$ of the randomized algorithm $\alg{A}$ over the input $x$ and uniformly randomness.
The min-entropy of a random variable $\mathbf{X}$ is $\minentropy(\mathbf{X}) = -\log\max_{x\in\mathbf{X}}\Prob{\mathbf{X} = x}$, and intuitively it measures the best chance to predict $\mathbf{X}$ (by a computationally unbounded algorithm).

\paragraph*{Negligible functions}\label{par:negl}
We denote by $\secpar\in\NN$ the security parameter and we implicitly assume that every algorithm takes as input the security parameter. A function $\nu:\NN\rightarrow[0,1]$ is called \emph{negligible} in the security parameter $\secpar$ if it vanishes faster than the inverse of any polynomial in $\secpar$, \emph{i.e.}\ $\nu(\secpar)\in \bigO{1/p(\secpar)}$ for all positive polynomials $p(\secpar)$.
We sometimes write $\negl$ (resp.,\ $\poly$) to denote an unspecified negligible function (resp.,\ polynomial function) in the security parameter.

\subsection{Blockchain Protocol}\label{subsec:blockchain-def}
A blockchain $\chain$ is a sequence of public and immutable blocks $\block$ administrated by the network.
A blockchain protocol $\Pi$ over a chain $\chain$ is a tuple of four algorithms $(\kgen, \allowbreak \elect, \createBlock, \ver)$, executed by block-creators in order to create new blocks.
The key generation algorithm $\kgen$ allows block-creators to join the network and generate their public and secret keys.
We denote with $\cBlockcreators = \{\blockcreator_1, \blockcreator_2, \ldots\}$ and $(\pk_{\blockcreator_i}, \sk_{\blockcreator_i})$, the block-creators present in the system and their public and secret keys, respectively.
At each timestamp $t$, block-creators work to extend the blockchain by appending a new block $\block$.\footnote{The new block will be cryptographically linked to its predecessor. Usually, such link is created by including in the new block the hash of the previous one.}
The protocol is composed of two distinct phases: \emph{Election phase} and \emph{creation phase}.
The election phase starts at the beginning of each timestamp $t$.
Each block-creator executes $\elect$ to reach consensus on a \emph{leader} that will be in charge of creating and appending a new block.
Once the consensus is reached, $\elect$ outputs an identification string $\selval$ that publicly identifies the leader $\blockcreator^*$ (\emph{e.g.}, $\selval$ includes the leader's public key $\pk_{\blockcreator^*}$).
Then, the creation phase starts: $\blockcreator^*$ broadcasts a new block $\block$ (generated by executing $\createBlock$) and the identification string $\selval$ over the network in order to be verified.
The blockchain is extended by appending $\block$ if and only if both $\block$ and $\selval$ are valid, and $\block$ has been created by the leader $\blockcreator^*$ identified by the identification string $\selval$.

More formally, a blockchain protocol consists of the following four algorithms:
\begin{description}
  \item[$\kgen(\secparam)$:] The randomized key generation algorithm takes as input the security parameter and outputs a public and secret key $(\pk_\blockcreator, \sk_\blockcreator)$.
  \item[$\elect(\pk_\blockcreator, \sk_\blockcreator)$:] The randomized consensus algorithm takes as input a public and secret key pair $(\pk_\blockcreator,\sk_\blockcreator)$ and outputs an identification string $\selval$.
  \item[$\createBlock(\pk_\blockcreator, \sk_\blockcreator, \selval)$:] The randomized creation algorithm takes as input the public and secret key pair $(\pk_\blockcreator, \sk_\blockcreator)$, and an identification string $\selval$, and outputs a block $\block$.
  \item[$\ver(\block, \selval)$:] The deterministic verification algorithm takes as input a block $\block$ and an identification string $\selval$, and outputs a decision bit.
\end{description}
We assume that leaders act in a rational manner, broadcasting new blocks as soon as they are ready.
This is a common behavior presents in most of the existing blockchains.
Timestamps have a predefined time length that is the expected average time needed to run the protocol and produce new valid blocks.
To make the system live, any delay is automatically handled by the network (\emph{e.g.}, electing another leader).
Furthermore, in blockchain systems such as Bitcoin and Ethereum, block-creators compete to mine new blocks and earn the corresponding coinbase transaction.
Any voluntary delay would result in an economic loss.

We now define the correctness.
Intuitively, a blockchain protocol $\Pi$ is correct if an honest execution produces a valid block $\block$.
\begin{definition}[Correctness of blockchain protocol]
We say that a blockchain protocol $\Pi$ is correct if, for every $\secpar \in \NN$, set of block-creators $\cBlockcreators = \{\blockcreator_1, \blockcreator_2, \ldots\}$ with keys $\{(\pk_{\blockcreator_i},\sk_{\blockcreator_i}) \getsr \kgen(\secparam)\}_{i \in [|\cBlockcreators|]}$ we have:
\[
  \Prob{
    \begin{array}{l}
      \selval \getsr \elect(\pk_{\blockcreator_i},\sk_{\blockcreator_i}), \\
      \ver(\createBlock(\pk_{\blockcreator^*}, \sk_{\blockcreator^*}), \selval) = 1
    \end{array}
    } \geq 1 - \negl,
\]
where $\blockcreator^*$ is the elected block-creator identified by $\selval$.
\end{definition}

\begin{remark}
Network partitioning and asynchronous communication may interfere during the election and bring block-creators to have different consensus views, \emph{i.e.}, $\elect$ algorithm may return inconsistent identification strings $\selval$.
This ends in having multiple leaders that create new blocks, generating blockchain forks.
We implicitly assume that the system has a mechanism to handle and solve forks automatically.
\end{remark}

\begin{remark}\label{rmrk:defs}
Our definition focuses on blockchain protocols that are composed of two distinct phases: Election phase ($\elect$ algorithm) and creation phase ($\createBlock$ algorithm).
An example of systems that lie in this category are permissioned blockchains.
In a permissioned setting, block-creators could jointly elect a leader (\emph{e.g.}, by running a cooperative consensus algorithm) that will be in charge of generating a new block $\block$.
Even the well-known permissionless blockchains Bitcoin and Ethereum fall in this category.
In these systems, the block creation works as a form of \emph{self-election}.
Indeed, block-creators are required to locally solve the PoW in order to produce a valid block.
The coinbase transaction of a newly mined block uniquely identifies the creator.
Hence, according to our definition, the election and creation phases collapse into a single one, \emph{i.e.}, the mined block can be seen as an identification string (\emph{e.g.}, $\selval = \block$), and the creation algorithm $\createBlock$ is just the identity function.
\end{remark}

We are interested in blockchain protocols that are unpredictable.
Unpredictability refers to the inability to predict the output of the election phase, \emph{i.e.}, the identification string $\selval$.

\begin{definition}[Unpredictability]\label{def:unpredict}
A Blockchain Protocol $\Pi$ is unpredictable if $\minentropy(\mathbf{X}) \geq \omega(\log(\secpar))$,
where $\mathbf{X}$ is a random variable that represents the distribution of the identification strings $\selval$ (output by $\elect$).
\end{definition}

\begin{remark}\label{rmrk:unpredictability}
Several blockchain systems are considered unpredictable.
For example, as described in Remark~\ref{rmrk:defs}, Bitcoin's identification string $\selval$ corresponds to the next candidate block $\block$.
Each block contains several elements such as: $32$ bit nonce used to randomize the output of the PoW, ECDSA signatures, Merkle root, extra-nonce, etc.
These elements have a non-trivial amount of entropy and make hard the prediction of the next block.
We also emphasize that, in some cases, it is straightforward to make a blockchain system unpredictable.
For example, in a blockchain that deploys a cooperative consensus algorithm (\emph{e.g.}, Ripple~\cite{rippleconsensus}), block-creators can jointly agree on randomness $r$ (\emph{e.g.}, by using coin tossing~\cite{AlonO16} or other multi party computation techniques) with at least $\omega(\log(\secpar))$ bits of min-entropy during the consensus phase.
\end{remark}

\subsection{Provable Data Possession}\label{subsec:pdp}
A Provable Data Possession scheme (PDP) $\Pi=(\kgen, \tagblock, \allowbreak \genchal,\allowbreak \genproof,\allowbreak \checkproof)$ allows a user to check the integrity of a file $\file = \{\bfile_1, \ldots, \bfile_n\}$ stored in a remote untrusted server.
The user computes the file \emph{fingerprint} (by using its public and secret key $(\pk, \sk)$ generated by $\kgen$) that consists in tagging each file chunk $\bfile_i$ using the tagging algorithm $\tagblock$.
Then, both the file $\file$ and the tags $\{\btag_i\}_{i \in [n]}$ are outsourced to the untrusted server.
At any moment, the user can issue a challenge $\chal$ (generated through the challenge generation algorithm $\genchal$) to the server in order to audit its data.
We assume that the challenge generation algorithm takes in input an integer $\fileproofNum$ and an index space $\cIndex \subseteq  [n]$ such that $\fileproofNum \leq |\cIndex|$, and returns $\chal$ that, among other things, includes a set of $\fileproofNum$ distinct indexes $\cIndex_{\chal}$ sampled at random from $\cIndex$.
The server uses both the file and the related tags to run $\genproof$ and generate a proof of possession $\pi$.
The latter is sent to the user that verifies it by running the verification algorithm $\checkproof$.
In this work, we are interested in publicly verifiable PDP, \emph{i.e.}, the verification process does not involve the user's secret key $\sk$.
This permits them to delegate a third party for verification.

Formally, a publicly verifiable PDP scheme for a file $\file = \{\bfile_1, \ldots, \allowbreak  \bfile_n\}$ is composed by the following algorithms:
\begin{description}
\item[$\keygen(\secparam)$:] The randomized key generation algorithm takes as input the security parameter and outputs the public key $\pk$ and the secret key $\sk$.
\item[$\tagblock(\pk,\sk,\bfile_i)$:] The randomized tagging algorithm takes as input a public key $\pk$, a secret key $\sk$, and a file chunk $\bfile_i \in \file$, and outputs a tag $\btag_i$.
\item[$\genchal(\fileproofNum, \cIndex)$:] The randomized challenge algorithm takes as an integer $\fileproofNum \in \NN$ and an index space $\cIndex \subseteq [n]$, and outputs a challenge $\chal$ defined over $\fileproofNum$ distinct chunk indexes $\cIndex_{\chal}$ sampled at random from $\cIndex$
\item[$\genproof(\pk,\chal,\{\bfile_i\}_{i\in\cIndex_\chal},\{\btag_i\}_{i \in \cIndex_\chal})$:] The deterministic prove algorithm takes as input a public key $\pk$, a challenge $\chal$ defined over a set of $\fileproofNum$ chunk indexes $\cIndex_\chal$, a set of file chunks $\{\bfile_i\}_{i\in\cIndex_\chal}$, and a set of tags $\{\btag_i\}_{i\in\cIndex_\chal}$, and outputs a proof $\pi$.
\item[$\checkproof(\pk,\chal,\pi)$:] The deterministic verification algorithm takes as input a public key $\pk$, a challenge $\chal$, and a proof $\pi$, and outputs a decision bit.
\end{description}
\noindent A PDP scheme is correct if honestly generated proofs verify correctly.
\begin{definition}[Correcteness of PDP]\label{def:correctnessPDP}
A PDP scheme $\Pi = (\kgen, \allowbreak \tagblock, \allowbreak\genchal, \genproof, \checkproof)$ is correct if $\forall \cIndex \subseteq [n]$, $\forall \file = \{\bfile_1, \allowbreak \ldots, \bfile_n\}$, $\forall \fileproofNum \in \NN$ such that $\bfile_i \in \bin^*$ and $\fileproofNum \leq |\cIndex|$, we have:
\[
\Prob{
  \begin{array}{l}
    \pi = \genproof(\pk, \chal, \mathcal{F}, \mathcal{T}), \\
    \checkproof(\pk, \chal, \pi) = 1
  \end{array}
  } = 1,
\]
where $(\pk, \sk) \getsr \kgen(\secparam)$, $\chal \getsr \genchal(\fileproofNum, \cIndex)$, $\cIndex_\chal$ are the $\fileproofNum$ chunk indexes determined by $\chal$, $\mathcal{F}=\{\bfile_i\}_{i\in\cIndex_\chal}$, and $\mathcal{T} = \{\btag_i \getsr \tagblock(\pk, \sk, \bfile_i) \in \cIndex_\chal, \}_{i\in\cIndex_\chal}$.
\end{definition}

\noindent As for security, it must be infeasible to generate a valid proof of possession $\pi$ without knowing the file chunks specified in the challenge.
\begin{definition}[Security of PDP]\label{def:securityPDP}
A PDP scheme $\Pi = (\kgen, \allowbreak \tagblock,\allowbreak \genchal, \allowbreak\genproof, \checkproof)$ is secure if, for every file $\file=\{\bfile_1, \ldots, \bfile_n\}$, every index space $\cIndex\subseteq[n]$, every $\fileproofNum \leq n$, and every PPT adversary $\adversary$, the probability that $\adversary$ wins the game $\gamePDP_{\Pi,\file,\adversary}(\secpar,\fileproofNum, \cIndex)$ is negligibly close to the probability that the challenger can extract the chunks $\{\bfile_{i}\}_{i \in \cIndex_\chal}$ by means of a knowledge extractor $\extract$.
The game $\gamePDP_{\Pi,\file,\adversary}(\secpar,\fileproofNum, \cIndex)$ is defined in the following way:
\begin{enumerate}
  \item The challenger runs $(\pk, \sk) \getsr \kgen(\secparam)$ and sends $\pk$ to the adversary $\adversary$.
  \item $\adversary$ issues oracle queries to the oracle $\tagblock$. On input $\bfile_i \in \file$ the oracle returns the tag $\btag_i \getsr \tagblock(\pk, \sk, \bfile_i)$.
  \item The challenger sample a randomness $r\getsr\bin^*$ and send it to $\adversary$.
  The adversary is required to provide a valid proof of possession $\pi$ with respect to the challenge $\chal = \genchal(\fileproofNum,\cIndex;\allowbreak r)$ defined over the $\fileproofNum$ indexes $\cIndex_\chal \subseteq \cIndex$.
  \item The adversary outputs a proof $\pi$.
  \item If $\checkproof(\pk, \chal, \pi) = 1$ where $\chal = \genchal(\fileproofNum, \cIndex;\allowbreak r)$, output $1$; otherwise output $0$.
\end{enumerate}
\end{definition}

\begin{remark}
The PDP definitions used in this work differs from the ones proposed by Ateniese et al.~\cite{ateniese2007provable} in how the challenges are generated.
We assume the existence of a randomized algorithm $\genchal$ that produces challenges on a portion of the file.
This main difference affects our security definition.
Indeed, similarly to~\cite[Section 4]{ateniese2009proofs}, the challenger does not directly choose $\chal$, but it outputs the randomness $r$ for the challenge algorithm $\genchal$ that will randomly sample $\fileproofNum$ chunks from the index space $\cIndex$.
Note that every secure PDP scheme under the definition of~\cite{ateniese2007provable}, is also secure under our Definition~\ref{def:securityPDP}.
This is because in~\cite{ateniese2007provable}, the challenge distribution is arbitrary.
\end{remark}

\begin{remark}
PDP and PoR~\cite{shacham2008compact} are related primitives.
The former guarantees that the challenged chunks are correctly stored (or known) by the server, while the latter additionally guarantees that the entire file can be retrieved from a set of proofs.
It is possible to show that retrievability can be added to PDP by applying an erasure code to the file $\file$ before uploading it~\cite{ateniese2011remoteJournal}. Thus, we implemented \systemname\ with PDP and left the use of erasure codes optional.
\end{remark}

\subsection{Signature Schemes}\label{subsec:sig}
A signature scheme is made of the following efficient algorithms.
\begin{description}
  \item[$\kgen(\secparam)$:] The randomized key generation algorithm takes the security parameter and  outputs a public and a secret key $(\pk, \sk)$.

  \item[$\sign(\sk, \msg)$:] The randomized signing algorithm takes as input the secret key $\sk$ and a message $\msg\in\cMSG$, and produces a signature $\sigma$.

  \item[$\ver(\pk, \msg,\sigma)$:] The deterministic verification algorithm takes as input the public key $\pk$, a message $\msg$, and a signature $\sigma$, and it returns a decision bit.
\end{description}

A signature scheme must satisfy two properties: $1)$ honestly generated signatures verify correctly and, $2)$ it is infeasible to compute valid signatures for new fresh messages without knowing the respective secret key $\sk$.
\ifFULL
\begin{definition}[Correctness of signatures] \label{def:Sigcorrectness}
    A signature scheme $\Pi=(\kgen, \allowbreak \sign, \ver)$ with message space $\cMSG$ is correct if $\forall \msg \in \cMSG, \forall (\pk, \sk) \getsr \kgen(\secparam)$
    the following holds:
    \[
        \Prob{\ver(\pk, \msg,\sign(\sk, \msg)) = 1} = 1.
    \]
\end{definition}

\begin{definition}[Unforgeability of signatures] \label{def:SigUnforgeability}
    A signature scheme $\Pi=(\kgen, \allowbreak \sign, \ver)$ is existentially unforgeable under chosen-message attacks (EUF-CMA) if for all PPT adversaries $\adversary$:
    \[
        \Prob{\Sigeufgame_{\Pi,\adversary}(\secpar) = 1} \leq \negl,
    \]
    where $\Sigeufgame_{\Pi,\adversary}(\secpar)$ is the following experiment:
    \begin{enumerate}
        \item $(\pk, \sk) \getsr \kgen(\secparam)$.
        \item $(\msg, \sigma) \getsr \adversary^{\sign(\sk, \cdot)}(\secparam, \pk)$
        \item If $\msg$ has not been queried to oracle $\sign(\sk, \cdot)$, and $\ver(\pk,\msg, \allowbreak \sigma) = 1$, output $1$; otherwise output $0$.
    \end{enumerate}
\end{definition}

\fi

\section{Assumptions}
\label{sec:Assumptions}
\systemname\ allows users to store their files in a decentralized way.
For the sake of clarity, we introduce the system assuming the presence of a single user with a single file.
Erasure code may be pre-applied in order to add redundancy and guarantees a user to retrieve the file even if a part of the chunks are lost or corrupted. We assume the file size is too large (\emph{e.g.}, Petabytes) to be stored by an individual node.
For this reason, the file is divided into smaller portions that are distributed to each node.
The distribution is performed by a third party $\dealer$, called the \emph{dealer}.
In Section~\ref{sec:discussion} we discuss how to handle multiple files and decentralize the distribution.
Additionally, we make the following assumptions:

\paragraph*{Rational nodes}\label{par:rational}
We assume that the majority of nodes (block-creators and storage-nodes) are rational and act in an economically rational manner.
Rewards are widely used in several blockchain systems to incentivize honest behavior.
In both Bitcoin and Ethereum, peers earn coins for the service provided to the network.
Incentivization can also be obtained in other ways.
As an example, in a permissioned scenario, the network is composed of authorized nodes.
It is reasonable to assume that, in this case, nodes have a contractual agreement with an authority that must be fulfilled.

\paragraph*{Secrecy of private keys}\label{par:privkey}
Each node possesses a public and secret key $(\pk, \sk)$.
We assume $\sk$ is kept secret, and thus not outsourced to an external party.
In permissionless systems such as Bitcoin or Ethereum, the node's secret key is used to sign transactions and spend their rewards.
Hence, a node that reveals its secret key $\sk$ would end in exposing its wallet.
On the other hand, in a permissioned setting, nodes must be authorized and $\sk$ allows them to authenticate in the system.
Revealing $\sk$ would allow an attacker to maliciously act in its name and breaks any existing contractual agreement.

\paragraph*{Network latency}\label{par:network}
We assume that communicating through the network requires a significant amount of time.
If a node desires to execute a computational task as fast as it can, it will be most likely faster if it minimizes the network communication by computing the task locally.
Outsourcing the computation to an external party will add a significant delay due to the multiple network hops necessary to exchange inputs and outputs for the task.
As we will describe in the Section~\ref{sec:discussion} this is fundamental in order to incentivize storage-nodes to locally store the file chunks and make \systemname\ resilient to outsourcing.

\section{\systemname}\label{sec:system-defs}
\systemname\ defines a new technique to add storage capabilities to every blockchain system (Definition~\ref{subsec:blockchain-def}).
The network is composed of two types of nodes: Block-creators and storage-nodes.
Block-creators execute the standard protocol of the underlying blockchain while storage-nodes are entitled to store a portion of the file.
\systemname\ requires cooperation between these two categories in order to create a new block.
A block-creator that wants to extends the blockchain with a new block $\block$ must challenge a subset of $\storagenodeNum$ storage-nodes and retrieve at least $\ticketproofnum$ proofs of possession $\{\pi_i\}_{i \in [\ticketproofnum]}$.
For the sake of clarity, we introduce the system assuming the presence of a single user with a single file and the distribution of the file chunks is performed by a third party $\dealer$, called the \emph{dealer}.
In Section~\ref{sec:discussion} we will discuss how to handle multiple files, decentralize the distribution, and other technical details required to correctly and securely instantiate the \systemname\ in a real scenario.

Formally, \systemname\ is composed of eight algorithms $(\bckgen, \allowbreak \snkgen, \setup, \allowbreak\getFrag, \elect, \prove, \allowbreak \createBlock, \allowbreak\ver)$ and consists of four distinct phases: \emph{Join}, \emph{distribution}, \emph{election}, and \emph{block creation}.

\paragraph*{Join} As in every blockchain system, nodes are free to join the network.
They only need to generate a valid key pair.
Since there are two categories of nodes, \systemname\ has two distinct key generation algorithms: $\bckgen$ generates the block-creators' keys while $\snkgen$ generates the storage-nodes' ones.
The storage-nodes' keys are valid keys of an arbitrary signature scheme (\emph{e.g.,} the same used by the underlying blockchain system).
We denote with $\cStoragenode = \{\storagenode_1, \storagenode_2, \ldots\}$ and $\{(\pk_{\storagenode_i}, \sk_{\storagenode_i})\}_{i \in [|\cStoragenode|]}$, the storage-nodes present in the system and their public and secret keys, respectively.

\begin{figure}[!t]
  \centering
  \includegraphics[scale=0.40]{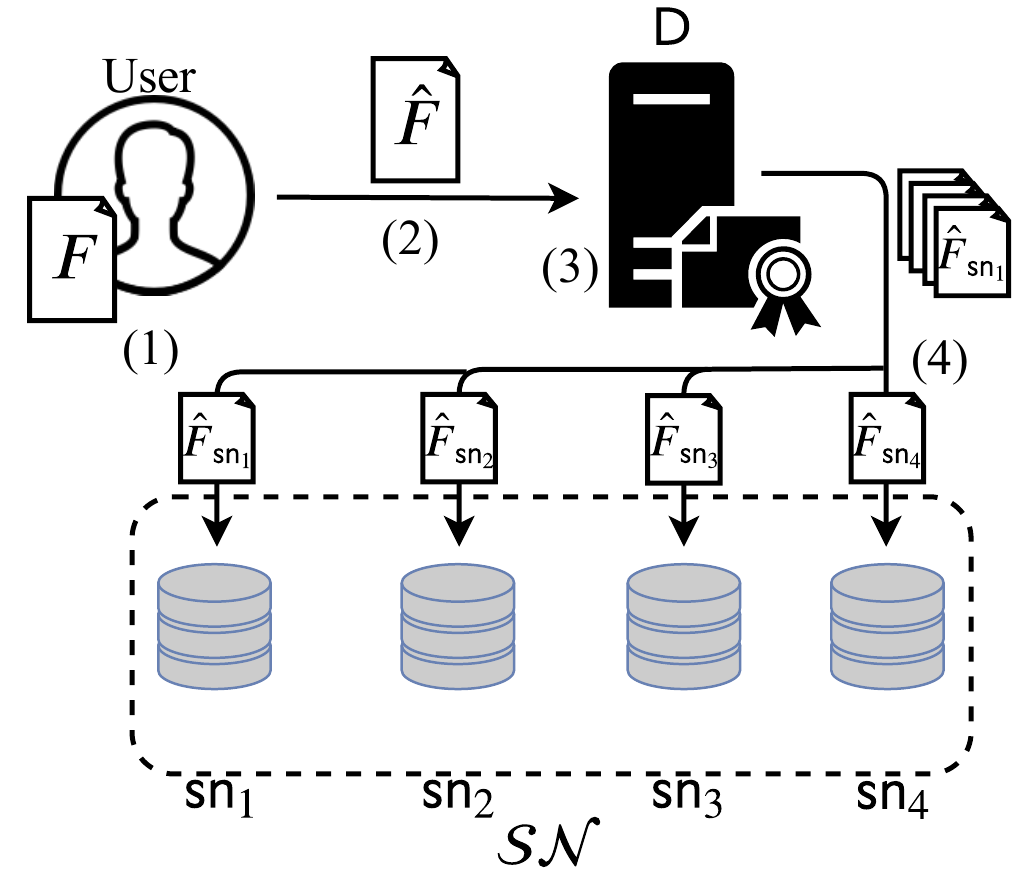}
  \caption{
  An example of file distribution with four storage-nodes $\cStoragenode = \{\storagenode_1, \storagenode_2, \storagenode_3, \storagenode_4\}$.
  $(1)$ The user, holding a file $\file$, executes $\setup$ and computes the file public key $\pk_{\encFile}$ and the encoding $\encFile$.
  $(2)$ The user sends the encoded file $\encFile$ to the dealer $\dealer$.
  $(3)$ $\dealer$ computes the file subsets $\{\subFile_{\storagenode_1}, \subFile_{\storagenode_2}, \subFile_{\storagenode_3}, \subFile_{\storagenode_4}\}$ by running $\getFrag$.
  $(4)$ $\dealer$ sends to each storage-node $\storagenode_i$ the respective file subset $\subFile_{\storagenode_i}$ to store.
  }
  \label{fig:distribution}
\end{figure}

\paragraph*{Distribution}
The distribution phase starts with a user that wants to store a file $\file = \{\bfile_1, \ldots, \bfile_n\}$.
It executes $\setup$ that outputs an encoding $\encFile = \{\encbFile_1, \ldots, \encbFile_n\}$ of $\file$ and the file public key $\pk_{\encFile}$ (\emph{i.e.}, file identifier).\footnote{
Such encoding may include the application of an erasure code on the file in order to add redundancy and guarantees the user to retrieve the file even if a part of the chunks are lost or corrupted.}
Among other things, the encoding $\encFile$ contains a set of tags $\{\btag_i\}_{i \in [n]}$, computed by a publicly verifiable PDP scheme.
The tags will allow storage-nodes to prove the file is correctly stored.
The user publishes $\pk_{\encFile}$ to announce its intention to store the file (see Section~\ref{par:pubparameters} for more details).
Then, it sends $\encFile$ to the dealer $\dealer$ whose job is to distribute to each storage-node $\storagenode$ a subset $\subFile_{\storagenode}\subseteq\encFile$ (computed by $\getFrag$) of $\storageblockNum$ chunks.
An example of distribution is depicted in Figure~\ref{fig:distribution}.

\begin{figure}[!t]
    \centering
      \includegraphics[width=0.7\columnwidth]{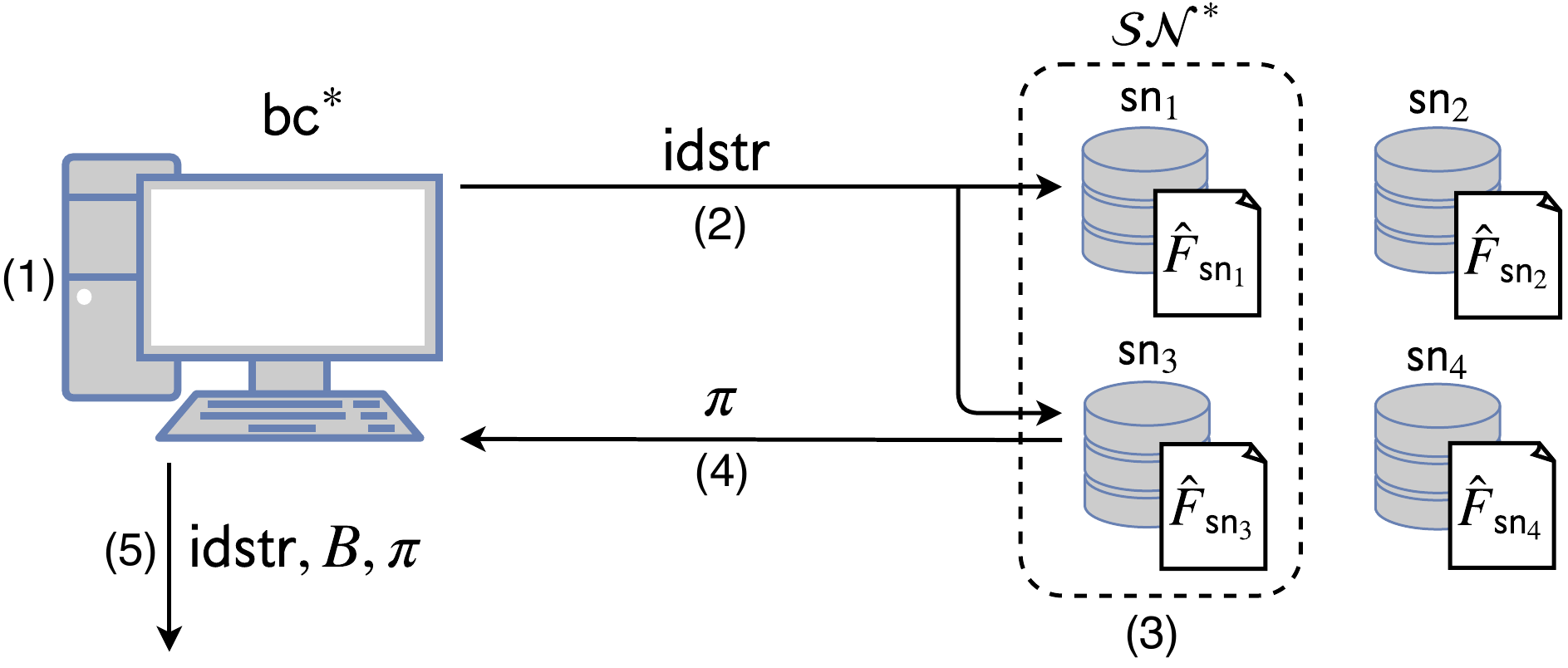}
  \caption{Creation phase execution with four storage-nodes $\cStoragenode=\{\storagenode_1, \storagenode_2, \storagenode_3, \storagenode_4\}$, $\storagenodeNum=2$, and $\ticketproofnum=1$.
  $(1)$ The network elects a leader $\blockcreator^*$ and a set of $\storagenodeNum$ storage-nodes $\cStoragenode^* = \{\storagenode_1, \storagenode_3\}$ by running $\elect$.
  $(2)$ The leader $\blockcreator^*$ sends the identification string $\selval$ to each storage-node $\storagenode^*_i \in \cStoragenode^*$.
  $(3)$ Each $\storagenode^*_i \in \cStoragenode^*$ executes $\prove$ and computes the proof $\pi$.
  $(4)$ $\storagenode^*_3$ (the fastest storage-node) sends $\pi$ to the leader $\blockcreator^*$.
  $(5)$ $\blockcreator^*$ broadcasts $\selval$, a new block $\block$, and $\pi$ for verification.
  }
  \label{fig:chain_extension}
\end{figure}

\paragraph*{Election} Once the distribution phase is completed, the election phase starts.
At each timestamp, block-creators try to reach consensus on a leader $\blockcreator^*$.
This is achieved by executing the election algorithm $\elect$ that outputs an identification string $\selval$ that identifies $\blockcreator^*$.
Furthermore, $\storagenodeNum$ distinct storage-nodes $\cStoragenode^*$ are elected.
Each elected storage-node is invited to prove the possession of its file portion by using the PDP scheme.

\paragraph*{Creation} The leader $\blockcreator^*$ sends to each elected storage-node $\storagenode^*_i \in \cStoragenode^*$ the identification string $\selval$ to prove that it is the leader for the current timestamp.
Then, each storage-node $\storagenode^*_i \in \cStoragenode^*$ sends back a proof of possession $\pi_i$ (computed by executing $\prove$) with respect to a challenge $\chal^*_i$ defined over $\fileproofNum$ chunk indexes $\cIndex_{\chal^*_i}$. The challenge $\chal^*_i$ is randomly generated by hashing the identification string $\selval$ and the storage-node's public key $\pk_{\storagenode^*_i}$.
We require the storage-node to sign $\pi_i$ with its secret key in order to attest the source of the proof.
The leader $\blockcreator^*$ broadcasts the identification string $\selval$, a new block $\block$ (generated by $\createBlock$), and the first received $\ticketproofnum$ proofs $\{\pi_i\}_{i \in [\ticketproofnum]}$.
Each node in the network executes the verification algorithm $\ver$ that checks the following: $i)$ $\selval$ is a valid identification string that identifies $\blockcreator^*$, $ii)$ $\block$ is valid block created by $\blockcreator^*$, $iii)$ $\{\pi_i\}_{i \in [\ticketproofnum]}$ are valid proofs of possession generated by $\ticketproofnum$ distinct elected storage-nodes.
If the verification succeeds, then the network extends the blockchain $\chain$ by appending the block $\block$.
Figure~\ref{fig:chain_extension} shows an example of creation phase execution.

We stress that the $\ticketproofnum$ valid proofs of possession are not included into the new block $\block$.
If the majority of block-creators is honest, then the creation of a new block implies that $\ticketproofnum\cdot \fileproofNum$ chunks have been proven correctly.

Below, we provide the formal instantiation of \systemname.

\begin{construction}\label{const:costruction}
Let $\chainsys$, $\pdp$, $\sig$ be a blockchain protocol, a public verifiable PDP scheme, and a signature scheme, respectively.\footnote{\systemname\ leverages a signature scheme to generate the key pair of storage-nodes. These keys are used to sign their proof of possessions and identify the storage-nodes in order to receive the rewards. The signature could be identical to the signature scheme used by the underlying blockchain.}
Let $\hash_1,\hash_2,\hash_3:\bin^*\rightarrow\bin^*$ be three distinct hash functions.
We build \systemname\ with parameters $(n, \storageblockNum, \storagenodeNum, \fileproofNum, \ticketproofnum)$ such that $\fileproofNum \leq \storageblockNum \leq n$ and $\ticketproofnum \leq \storagenodeNum \leq |\cStoragenode|$ in the following way:
\begin{description}
	\item[$\bckgen(\secparam)$:] The block-creator key generation algorithm, on input the security parameter, returns $(\pk_\blockcreator, \sk_\blockcreator) \getsr \allowbreak \kgen_\chainsys\allowbreak(\secparam)$.

	\item[$\snkgen(\secparam)$:] The storage-node key generation algorithm, on input the security parameter, returns $(\pk_\storagenode, \sk_\storagenode)\getsr\allowbreak\kgen_\sig(\secparam)$.

	\item[$\setup(\secparam, \file)$:] The setup algorithm, on input the security parameter and a file $\file = \{\bfile_1, \ldots, \bfile_n\}$, computes $(\pk,\sk)\allowbreak\getsr\keygen_\pdp\allowbreak(\secparam)$ and $\btag_i=\tagblock_\pdp(\pk,\allowbreak\sk,\bfile_i)$, for $i\in[n]$.
	Finally, it returns $\pk_{\encFile} = \pk$ and $\encFile = \{\encbFile_i = (\bfile_i, \btag_i)\}_{i\in[n]}$.

  \item[$\getFrag(\pk_{\encFile}, \encFile, \pk_{\storagenode})$:] The chunk distribution algorithm takes as input the file public key $\pk_{\encFile}$, the encoded file $\encFile = \{\encbFile_i = (\bfile_i, \btag_i)\}_{i\in[n]}$, and a storage-node public key $\pk_{\storagenode}$.
  Then, it initiliazes $\set{X}_\storagenode = \{\emptyset\}$, $\set{V} = \{1, \ldots, n\}$, $j = 0$ and it randomly samples without replacement $\storageblockNum$ file chunk indexes in the following way:\footnote{Other randomized algorithms can be used to sample without replacement, \emph{e.g.}, reservoir sampling~\cite{reservoir}.\label{footnote:reservoir}}
	\begin{itemize}
		\item Until $|\set{X}_{\storagenode}| < \storageblockNum$ then: Increment the counter $j = j +1$, compute $v_j = \hash_1(\pk_{\storagenode} || j)$, and if $v_j \not\in \set{X}_{\storagenode}$, then add $v_j$ to $\set{X}_{\storagenode}$.
	\end{itemize}
	Finally, it returns $\subFile_{\storagenode} = \{\encbFile_i = (\bfile_{v_i}, \btag_{v_i})\}_{v_i \in \set{X_\storagenode}}$.

  \item[$\elect(\pk_{\blockcreator},\sk_{\blockcreator})$:] The election algorithm, on input a block-creator public key and private key pair $(\pk_{\blockcreator},\sk_{\blockcreator})$, computes $\selval\allowbreak\getsr\elect_\chainsys(\pk_{\blockcreator},\sk_{\blockcreator})$.
	Then, it initializes $\cStoragenode^* = \{\emptyset\}$, $\set{V} = \cStoragenode$, $j = 0$ and it randomly samples without replacement $\storagenodeNum$ storage-nodes $\cStoragenode^*$ in the following way:\footnotemark[5]
	\begin{itemize}
		\item Until $|\cStoragenode^*| < \storagenodeNum$ then: Increment the counter $j = j +1$, compute $v_j = \hash_2(\selval || j)$ and, if $\storagenode_{v_j} \not\in \cStoragenode^*$, then add $\storagenode_{v_j}$ to $\cStoragenode^*$
	\end{itemize}
	Finally, it returns $(\selval,\cStoragenode^*)$.

  \item[$\prove(\pk_{\encFile}, \pk_{\storagenode}, \sk_{\storagenode}, \selval, \subFile_{\storagenode})$:] The prove algorithm takes as input the file public key $\pk_{\encFile}$, a storage-node public and secret key $(\pk_{\storagenode}, \sk_{\storagenode})$, the identification string $\selval$, and a subset of file chunks $\subFile_{\storagenode} = \{\encbFile_i = (\bfile_{i}, \btag_{i})\}_{i \in [m]}$.
	It generates a challenge by executing $\chal^*=\genchal_\pdp(\fileproofNum, \set{X}_{\storagenode};\hash_3(\pk_{\encFile} || \allowbreak \pk_{\storagenode} \allowbreak || \selval))$ where $\set{X}_\storagenode$ is the set of chunk indexes stored by $\storagenode$ (as described in $\getFrag$).
	Let $\cIndex_{\chal^*} \subseteq \set{X}_\storagenode$ be the set of $\fileproofNum$ chunk indexes determined by $\chal^*$.
	The algorithm computes $\sigma=\sign_\sig(\sk_{\storagenode}, \pi')$ where $\pi'=\genproof_\pdp(\pk_{\encFile}, \allowbreak \chal^*,\allowbreak \{\bfile_{j}\}_{j\in\cIndex_{\chal^*}},\allowbreak \{\btag_{j}\}_{j\in\cIndex_{\chal^*}})$.
	It returns a proof of possession $\pi = (\pi',\sigma)$.

  \item[$\createBlock(\pk_{\blockcreator}, \sk_{\blockcreator}, \selval)$:] The creation algorithm, on input a block-creator public and secret key pair $(\pk_{\blockcreator}, \sk_{\blockcreator})$ and an identification string $\selval$, runs $\block\getsr\createBlock_\chainsys\allowbreak(\pk_{\blockcreator}, \allowbreak \sk_{\blockcreator}, \selval)$ and returns the new block $\block$.

	\item[$\ver(\pk_{\encFile}, \{\pk_{\storagenode_i}\}_{i\in[\ticketproofnum]}, \{\pi_i\}_{i \in [\ticketproofnum]}, \selval, \block)$:] The algorithm takes as input the file public key $\pk_{\encFile}$, a set of storage-node public key $\{\pk_{\storagenode_i}\}_{i\in[\ticketproofnum]}$, a set of proof of possession $\{\pi_i = (\pi'_i, \sigma_i)\}_{i \in [\ticketproofnum]}$, an identification string $\selval$, and a block $\block$.
	The algorithm proceeds in the following way:
	\begin{itemize}
		\item Compute $\cStoragenode^*$ as done by the $\elect$ algorithm. If there exists a $\pk_{\storagenode_i}$ that belongs to a storage-node $\storagenode_i$ such that $\storagenode_i \not\in \cStoragenode^*$, then return $0$. Otherwise, for every $i \in [\ticketproofnum]$, compute the challenge $\chal^*_i=\genchal_\pdp(\fileproofNum, \set{X}_{\storagenode_i};\allowbreak \hash_3(\allowbreak\pk_{\encFile} || \pk_{\storagenode_i} ||\allowbreak \selval))$, $b^2_i=\ver_\sig(\pk_{\storagenode_i},\allowbreak \pi_i', \sigma_i)$, and $b^1_{i}=\checkproof_\pdp(\allowbreak\pk_{\encFile},\chal^*_i,\pi_i')$, where $\set{X}_{\storagenode_i}$ is the set of chunk indexes stored by $\storagenode_i$ (computed as in $\getFrag$ algorithm).
	\end{itemize}
	Finally, if $1 =\ver_\chainsys(\selval,\block)$ and $b^1_i = b_i^2 = 1$ for every $i \in [\ticketproofnum]$, the algorithm returns $1$; otherwise, it returns $0$.
\end{description}
\end{construction}

\noindent\systemname\ inherits from the PDP scheme the same security guarantee, \emph{i.e.}, every time the chain is extended then a set of $\ticketproofnum$ storage-nodes have provided $\ticketproofnum$ valid proofs of possession.
At high level, Theorem~\ref{thm:sec-system} means that the $\ticketproofnum$ storage-nodes, that have produced the valid proof of possessions, know the corresponding challenged chunks (either because they store the chunks or they know how to recompute them).\ifCONF\footnote{
The proof of Theorem~\ref{thm:sec-system} can be found in the full version of the paper \url{https://eprint.iacr.org/2019/1345}
.}\fi
\ifFULL
Below we establish the result, and the proof appears in appendix~\ref{sec:proof-thm}.
\fi

\begin{theorem}\label{thm:sec-system}
Let $t$ be a timestamp in which the blockchain has been extended.
Let $\file = \{\bfile_1, \ldots, \bfile_n\}$ and $\cStoragenode^*$ an arbitrary file and the elected storage-nodes at timestamp $t$, respectively.
If $\pdp$ is secure (Def.~\ref{def:securityPDP}), then \systemname\ guarantees (in the random oracle model) that there exists a set of $\ticketproofnum$ storage-nodes $\{\storagenode_i\}_{i \in [\ticketproofnum]} \subseteq \cStoragenode^*$ such that each $\storagenode^*_i$ has generated a proof of possession $\pi_i$ with probability negligibly close to the probability that the user can extract the challenged $\fileproofNum$ file chunks $\{\bfile_j\}_{j \in \cIndex_{\chal^*_i}}$ by means of a knowledge extractor $\extract$, where $\cIndex_{\chal^*_i}$ are the indexes contained in the challenge $\chal^*_i$ of storage-nodes $\storagenode^*_i$.
\end{theorem}

\subsection{Technical details of \systemname} \label{sec:discussion}

\paragraph*{Publishing the parameters}\label{par:pubparameters}
\systemname\ relies on two public parameters that the entire network must know: $1)$ An ordered list $\cStoragenode$ that contains the storage-nodes present in the system and, $2)$ the public key $\pk_{\encFile}$ of the file.
The ordered list $\cStoragenode$ is essential in order to check that the proofs of possession $\{\pi_i\}_{i \in [\ticketproofnum]}$ come only from elected storage-nodes $\storagenode^*_i \in \cStoragenode^*$.
On the other hand, $\pk_{\encFile}$ is needed to validate the proofs $\{\pi_i\}_{i \in [\ticketproofnum]}$.
\systemname\ introduces two types of transactions and leverages the blockchain to publish $\cStoragenode$ and $\pk_{\encFile}$.
The first, called \emph{join} transaction $\jointrans$, allows storage-nodes to join the network and publish their public keys. All the transactions recorded on the blockchain compose the ordered list $\cStoragenode$.
The second is called \emph{store} transaction $\storetrans$ and allows the user to publish the file public key $\pk_{\encFile}$.

\paragraph*{Financial model}\label{par:financialmodel}
In order to block denial of service attacks, \systemname\ charges the user to pay storage fees proportional to the time the file is stored.
Suppose the user intends to store the file for $t$ consecutive timestamps, then it must include $t\alpha$ coins in the store transaction $\storetrans$.
These coins will be gradually delivered to the fastest storage-nodes of the next $t$ timestamps, \emph{i.e.}, for each timestamp, the leader includes into its proposed block $\ticketproofnum$ transactions $\{\trans_{\storagenode^*_i}\}_{i \in [\ticketproofnum]}$, each of which transfers $\frac{\alpha}{\ticketproofnum}$ coins from $\storetrans$ to one of the fastest $\ticketproofnum$ storage-nodes.
We stress that the payment can also be delivered in other ways.
For example, in a permissioned setting, storage-nodes can have some kind of off-chain contracts.

This financial model makes \systemname\ resilient against outsourcing.
To extend the blockchain, leaders are required to broadcast $\ticketproofnum$ proofs of possession along with the new block $\block$.
For this reason, a rational leader, that intends to broadcast a new block as fast it can (see Section~\ref{subsec:blockchain-def}), will wait only for the first $\ticketproofnum$ incoming proofs.
This incentivizes storage-nodes to keep their data locally stored.
Indeed, outsourcing a significant portion of data to a third party will end in an economic loss.
This is because the proof generation will require communication through the network making the storage-node significantly slower than the others (see the security assumptions in Section~\ref{sec:Assumptions}).
Moreover, note that storage-nodes can not predict in advance which will be the challenged chunk indexes since they are computed by hashing the identification string $\selval$ that has a non-trivial amount of entropy (Def.~\ref{def:unpredict}).
We stress that \systemname\ is outsourcing free only if $\frac{\ticketproofnum}{\storagenodeNum}$ is small enough.
The higher the value (\emph{e.g.}, $\frac{\ticketproofnum}{\storagenodeNum} \approx 1$), the lower the storage-node competition: A large portion of storage-nodes may decide to outsource the data since a high number of proofs are required during the creation phase.
On the other hand, the lower the value (\emph{e.g.}, $\frac{\ticketproofnum}{\storagenodeNum} \approx 0$), the higher the competition: The leader will collect only a limited number of proofs. Any delay significantly decreases the probability of getting the reward.
Lastly, by requiring users to pay storage-nodes, \systemname\ makes any collusion strategy worthless.
Indeed, even if the user and a set of storage-nodes collude, their expected reward will be negative.
This discourages any collusion and mitigates attacks such that the one described in Section~\ref{subsec:Untrusteduser}.

To deploy this financial model, the block must be created after receiving the proofs of possession, \emph{i.e.}, $\createBlock$ is executed after the election algorithm $\elect$.
Unfortunately, this is not the case of the existing systems Bitcoin and Ethereum.
As described in Remark~\ref{rmrk:defs}, in these systems, the creation of the block is a form of self-selection (\emph{i.e.}, $\elect$ and $\createBlock$ collapse into a single algorithm).
This does not permit the leaders to include the transaction $\{\trans_{\storagenode^*_i}\}_{i \in [\ticketproofnum]}$ into the new block since it would require to recompute the PoW.
To overcome this problem, in this kind of systems, storage-nodes are paid in the next timestamp.
Before proposing the block to the network, the leader must add into the transaction pool $\{\trans_{\storagenode^*_i}\}_{i \in [\ticketproofnum]}$ that pay the fastest $\ticketproofnum$ storage-nodes $\{\storagenode^*_i\}$.
After that, it broadcasts the block $\block$ along with the proofs $\{\pi_i\}_{i \in [\ticketproofnum]}$, generated by $\{\storagenode^*_i\}_{i \in [\ticketproofnum]}$.
The network will accept the block $\block$ by additionally checking that the transaction pool contains $\{\trans_{\storagenode^*_i}\}_{i \in [\ticketproofnum]}$.

\paragraph*{Recovering the file}\label{par:retrieve-file}
As discussed in~\ref{subsec:retrievability}, a malicious storage-node $\storagenode$ may choose to not return the stored file portion $\subFile_\storagenode$.
This problem can be solved by adopting the solution proposed by KopperCoin~\cite{KoppBK16}.
The user and the storage-node $\storagenode$ create a $2$-out-of-$2$ multisignature transaction $\trans$ that includes three amounts: a payment $\alpha$, a two collaterals $\beta$ (the client one) and $\gamma$ (the storage-node one).
The collaterals $\beta$ and $\gamma$ are a form of warranty to encourage the parties to act honestly.
Once the user receives $\subFile_\storagenode$ from the storage-node $\storagenode$, they unlock $\trans$ by signing a new transaction $\trans'$ that returns the collaterals to the respective parties and $\alpha$ is paid to $\storagenode$.
We stress that in smart contract enabled blockchains, other solutions such as fair exchange protocols~\cite{dziembowski2018fairswap} can be used.

\paragraph*{Multiple files}\label{par:multi-files}
\systemname\ can be easily extended in order to store $c$ files $\{\encFile_i\}_{i \in [c]}$ (with public keys $\{\pk_{\encFile_i}\}_{i \in [c]}$) of $c$ different users.
For each file $\encFile_i$, the dealer distributes to each storage-node $\storagenode$ a portion $\subFile^i_{\storagenode}$ of $\encFile_i$ composed of $\storageblockNum$ chunks. As before, the portion $\subFile^i_{\storagenode}$ for a storage-node $\storagenode$ is computed by running $\getFrag$ with input $(\pk_{\encFile_i}, \encFile_i, \pk_{\storagenode})$.
During the creation phase, each elected storage-node $\storagenode^*_i \in \cStoragenode^*$ is now required to provide to the leader $c$ different proofs of possession $\{\pi_{i,j}\}_{j \in [c]}$, one for each of the stored file portion $\subFile^j_{\storagenode^*_i}$.
Lastly, a new block is accepted by the network only if it is accompanied by $\ticketproofnum \cdot c$ PDP proofs $\{\pi_{i,j}\}_{i \in [\ticketproofnum], j \in [c]}$ computed by $\ticketproofnum$ elected storage-nodes.
Note that when multiple files are stored, the memory and computational power required to join the network increase proportional to the number of files stored.
Indeed, each storage-node needs to store $c\cdot\storageblockNum$ chunks and must compute $c$ independent proofs.
To decrease these requirements, the parameters $\storageblockNum, \fileproofNum, \ticketproofnum$ can be tuned in the following way:
\begin{itemize}
  \item $\storageblockNum$ (the number of chunks to store for each file) can be reduced (\emph{e.g.}, $\storageblockNum' = \frac{\storageblockNum}{c}$) in order to decrease the amount of free space required to join the network.
  \item $\fileproofNum$ (the number challenged chunks determined by $\chal$) can be reduced to speed up the proof generation.
  \item When $\fileproofNum$ decreases, the system checks the integrity of a smaller portion of the files.
  This can be counterbalanced by using the parallelism that \systemname\ provides: By increasing $\ticketproofnum$ (the number of proofs of possession required to propose a new block) it is possible to check the integrity of bigger portions of the files leveraging the proofs of possession of different storage-nodes.
  In Section~\ref{sec:implementation} we will show the performance of \systemname\ with different values $\ticketproofnum$.
\end{itemize}

\paragraph*{Files sizes}\label{par:file-sizes}
The sum of the sizes of files stored by \systemname\ must be large enough that it can not be stored by an individual storage-node (see Section~\ref{sec:Assumptions}).
A malicious storage-node, that stores all the files chunks, may sybil multiple identities and increase its chance to be elected.
If the sybil attack is performed an infinitely large number of times, the probability will tend to $1$.
This will guarantee the malicious node to get the reward at each timestamp and, at the same time, will disincentivize other storage-nodes to participate, decreasing the decentralization (and the security) of \systemname.
We stress that such assumption is reasonable to be true when the system is deployed on large scale.

\paragraph*{Decentralized dealer}\label{par:decentralized-dealer}
The dealer is just responsible for distributing the files across the network, allowing the user to go offline.
Hence, it can be easily decentralized by replicating the dealer on multiple servers.
Alternatively, it is possible to leverage a smart contract based blockchain to implement the dealer with a smart contract.

\paragraph*{New storage-nodes}\label{par:new-storage-nodes}
A storage-node could join the system after the file distribution has been completed.
The (decentralized) dealer will serve these new storage-nodes by sending them the chunks to store.
We stress that the dealer does not need to keep the file locally stored, indeed, it can simply retrieve the chunks from the storage-nodes already in the system.
Lastly, we emphasize that a storage-node could refuse to send the requested chunks back since a new node in the system decreases its chance to be elected.
To punish and disincentivize this behavior, the dealer marks the malicious node as faulty and excludes it from the system.

\section{Implementation and Evaluation}\label{sec:implementation}
We implemented a prototype to validate and demonstrate the technical feasibility of \systemname. We implemented the system as a decentralized application via smart contract logic~\cite{EthereumWhite}. Smart contracts allow us to build a functioning system without modifying the consensus mechanism and the block structure of a forked blockchain. The choice to implement a decentralized application is justified by two main reasons: a pragmatic and a technical one. The pragmatic one is related to the opportunity to reduce the complexity of the implementation while proving the functioning of the protocol.
The technical one is to be able to demonstrate that \systemname\ can be built on a smart contract enabled blockchain without touching the core protocol of the platform.
Naturally, as already described, \systemname\ can be implemented by modifying the protocol of an existing blockchain.

\ifFULL
\begin{figure}[!t]
 \centering
 \includegraphics[scale=0.5]{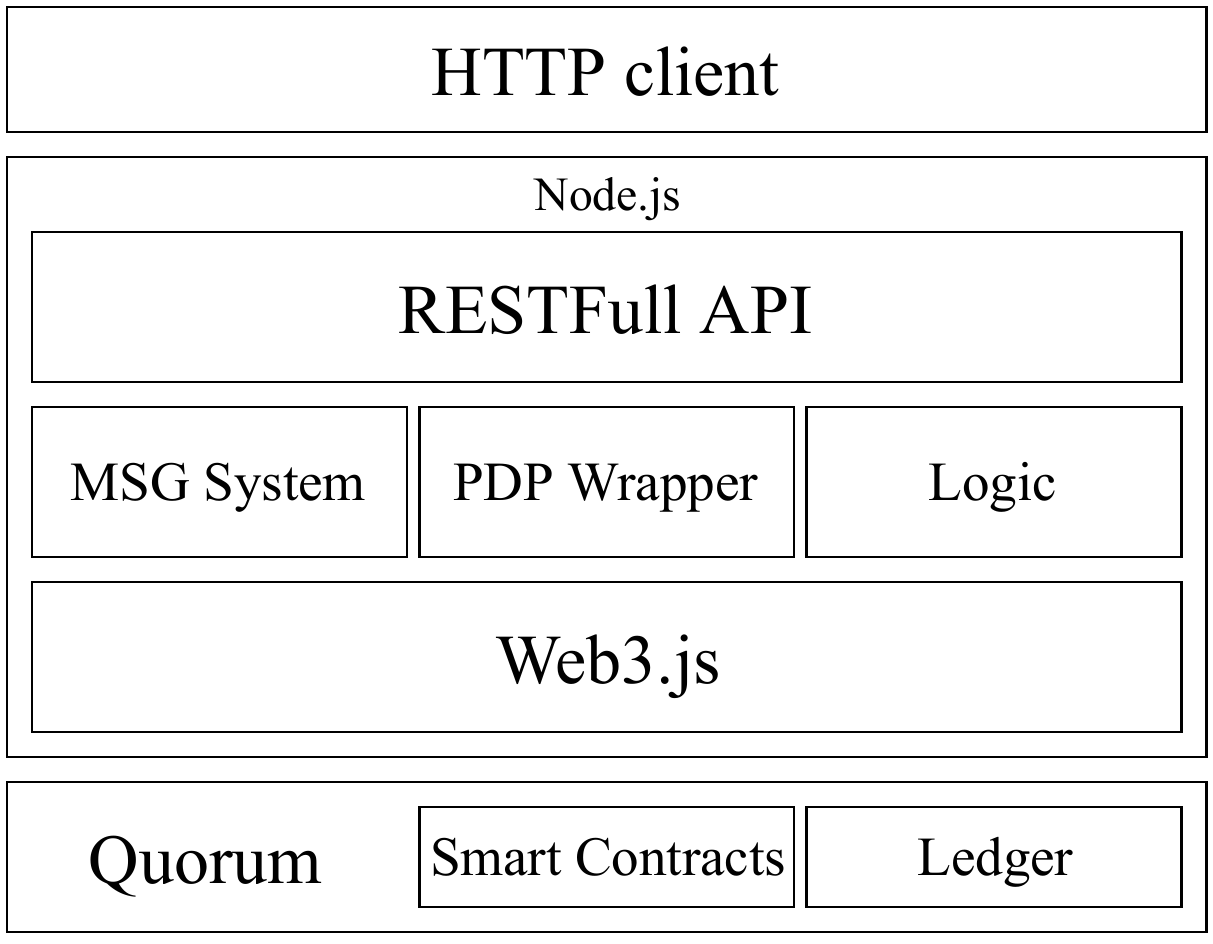}
 \caption{High level architecture.}
 \label{fig:implementation_arch}
\end{figure}
\fi

The client architecture \ifFULL(depicted in Figure \ref{fig:implementation_arch}) \fi
is structured with two main building blocks: a smart contract enabled blockchain and a server component.
The blockchain platform selected for our implementation is the Ethereum-based distributed ledger protocol Quorum~\cite{Quorum}. The version of Quorum used in our implementation is the 2.1.1, the geth version used is 1.73, the consensus mechanism configured in our system is RAFT~\cite{Ongaro2014InSO} and the smart contracts are written in Solidity 0.4.19. The server component is implemented in Node.Js 8.12.0 and has 5 main building blocks:
\ifFULL
\begin{itemize}
\fi
\ifCONF
\begin{enumerate*}[label={\arabic*})]
\fi
 \item A RESTFull API Layer for HTTPS client interactions;
 \item A PDP Wrapper connected to the PDP subroutine;
 \item A messaging system for off-chain communication;
 \item Web3.js (1.0.0-beta.36 with a custom patch to overcome some limitation on WebSocket Handling);
 \item Participants logic libraries to implement specific logic for each of the different roles, \emph{i.e.}, dealer, block-creator, and storage-node.
\ifFULL
\end{itemize}
\fi
\ifCONF
\end{enumerate*}
\fi
The PDP subroutine implements the publicly verifiable PDP scheme of Ateniese et al.~\cite{ateniese2007provable}, and it is based on an existing implementation called libpdp~\cite{libpdp}. We modified the libpdp library to implement the variant of the primitive that offers public verifiability.
The server component is the core logic of the different roles, and it is responsible for computing/generate proofs of possession using the library.
Additionally, it allows off-chain communication to send/receive proofs as well as the chunks to store.
In our implementation, the dealer (resp. the leader $\blockcreator^*$ of a fixed timestamp) executes a dedicated smart contract to compute the chunk indexes a storage-node must store (resp. to compute the storage-nodes $\cStoragenode^*$ elected in a particular timestamp).\footnote{The smart contract only computes the chunk indexes. The real chunks are sent off-chain.}
This makes the blockchain a public record enabling public auditing and transparency about the file distribution and the storage-node elections.

For the sake of efficiency, in our implementation, the election phase is simulated by an external party called \emph{oracle}.
The oracle beats the time for the network by communicating the start of new timestamps.
A new timestamp starts when the oracle executes the election phase, implemented by a smart contract, that on input a random seed $s$, selects the leader $\blockcreator^*$. The leader will then use the same seed $s$ to compute the elected storage-nodes $\cStoragenode^*$ and the challenge $\chal$.\footnote{According to the syntax provided in Section~\ref{subsec:blockchain-def}, the identification string $\selval$ of our implementation corresponds to the tuple $(\pk_{\blockcreator^*}, s)$.}
We stress that the oracle has been introduced only to reduce the complexity of the implementation.
The hash function used in our implementation is SHA-3.

\begin{figure}[!t]
 \centering
 \includegraphics[width=0.8\columnwidth]{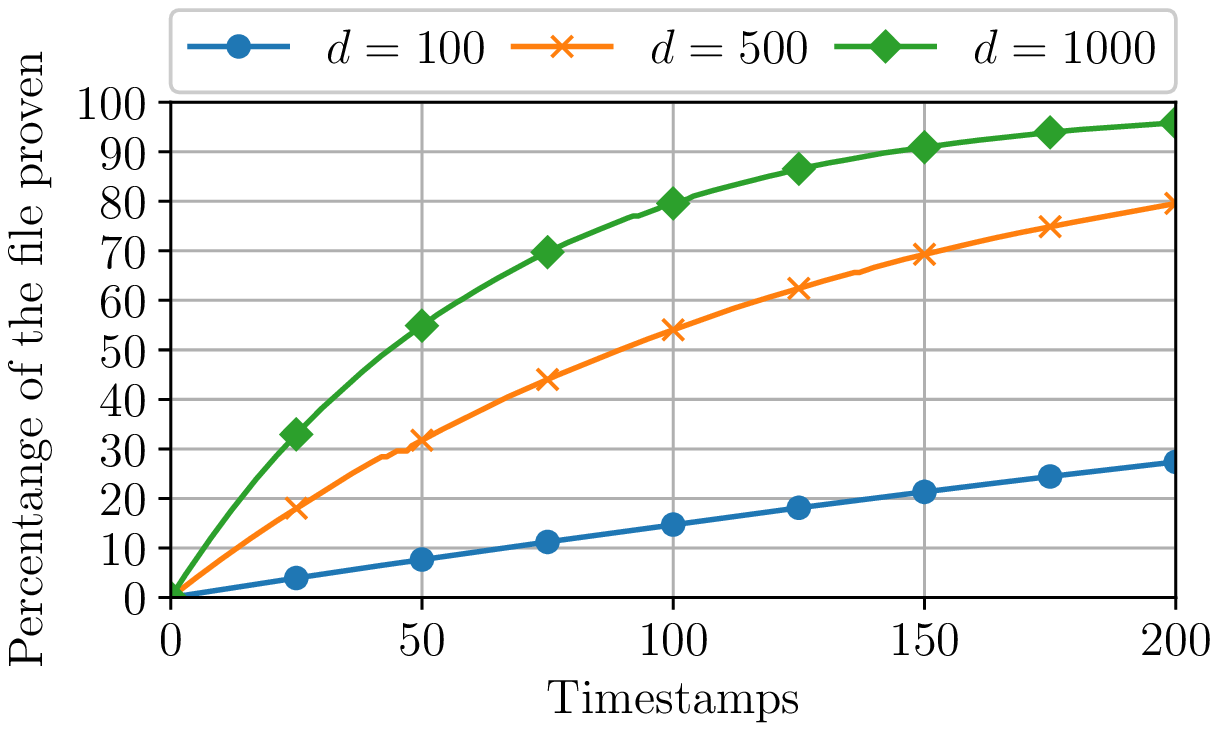}
 \caption{Percentage of the file proven over time.}
 \label{fig:accenture_test}
\end{figure}

\begin{figure}[!t]
 \centering
 \includegraphics[width=\columnwidth]{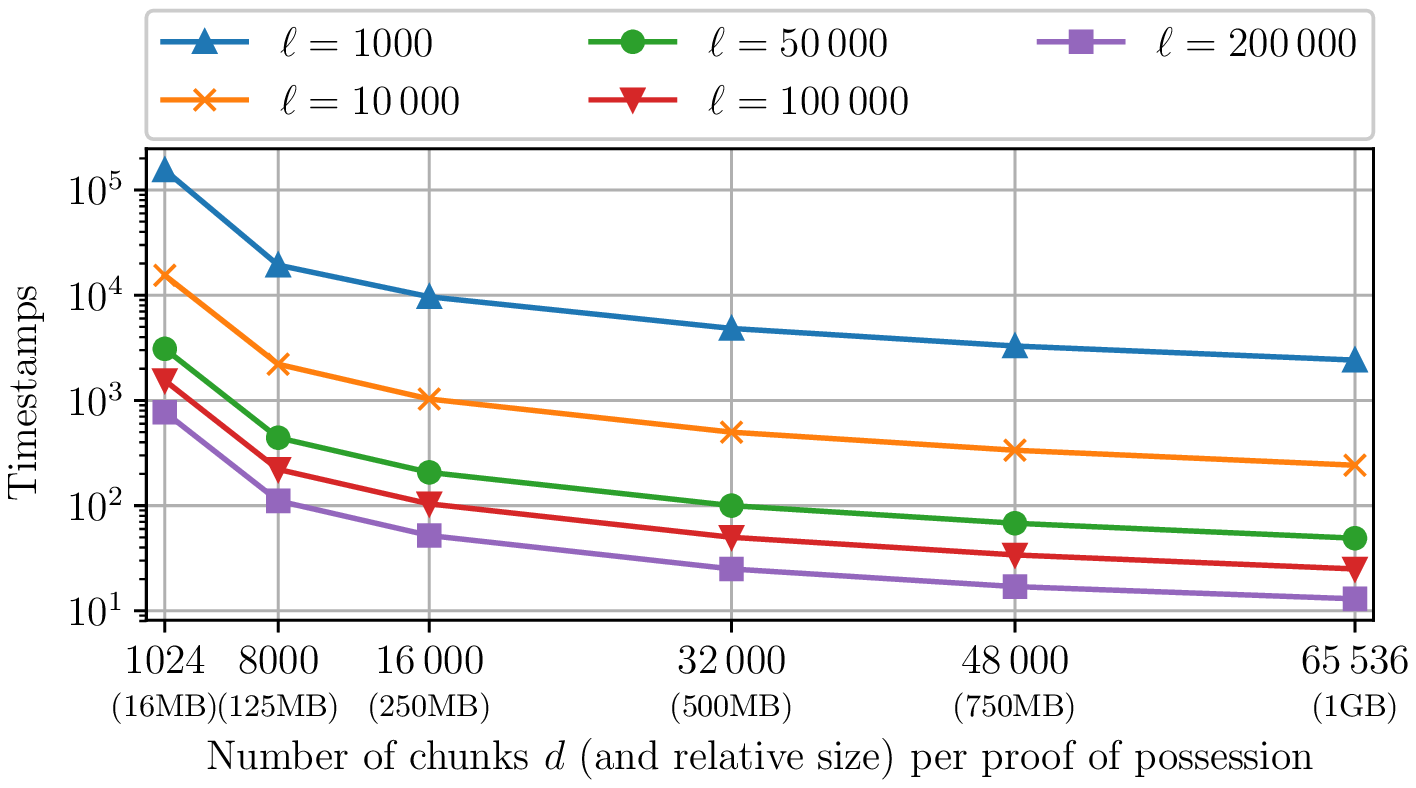}
 \caption{Number of timestamps needed to prove $90\%$ of the $1$ Petabyte file (composed of $n = \num{68719476736}$ chunks). The Y-axis (timestamps number) follows a logarithmic scale.}
 \label{fig:test_multiple_l}
\end{figure}

\paragraph*{Experimental setup} We deployed $7$ Quorum nodes as docker containers using Quorum Maker~\cite{quorummaker}. The host machine is an Amazon t3.2xlarge EC2 machine ($8$ vCPUs and $32$GB of RAM), running Ubuntu Xenial 16.04 amd64. Therefore, the ledger is only replicated $7$ times. The participants of the \systemname\ network do not have their own copy of the ledger.
The \systemname\ network has been deployed with a set of $10$ Amazon m5.12xlarge EC2 machines ($48$ cCPUs and $192$ GB of RAM), running Ubuntu Xenial 16.04 amd64.
Each virtual machine runs $100$ storage-nodes, $2$ block-creators and $2$ dealers as dockers containers.
One of the $10$ virtual machines runs the oracle, as a docker container, that elects the leader at each timestamp.
Overall, the \systemname\ network is composed of: $1000$ storage-nodes, $20$ block-creators, $10$ dealers, $1$ oracle.

To optimize the duration of the file-sharing process on our experiments, we used the docker bind mount process~\cite{bindmounts} instead of the originally implemented HTTP protocol. Each storage-node container is bound to a folder on the host machine where the dealer copies the right chunks.
This allows a dealer to successfully send the appropriate chunks to the right storage-nodes (hosted on the same machine) without going through the HTTP protocol.

\paragraph*{Experimental results}

We performed an experimental test and a simulation to evaluate the storage guarantees that \systemname\ provides with respect to a single file.
The results are identical even when the system stores multiple files.
The execution time of the protocol heavily depends on the PDP scheme used and the underlying blockchain (block creation time, number of storage-nodes, etc.).

We started with a performance test with a minimal instantiation of the system (small file and limited number of storage-nodes).
We measured the relation between the number of timestamps and the percentage of the file that is being processed.
In more detail, we executed our test by considering the following parameters: $1$GB file composed by $n=\num{65536}$ chunks (chunk size $16$KB), number of chunks to store by storage-node $\storageblockNum = \num{12500}$ ($\approx 19$\% of the entire file), number of proofs per timestamp $\ticketproofnum = 1$, number of elected storage-nodes per timestamp $\storagenodeNum = 10$.
Based on these parameters, we ran three experiments with different values of $\fileproofNum$ (number of chunks proven by each proof of possession): $100$, $500$, and $\num{1000}$ (approximately $0.15$\%, $0.8$\%, and $1.6$\% of the file).



Figure~\ref{fig:accenture_test} shows the results.
As we can see, the parameter $\fileproofNum$ has a significant impact on the percentage of the file chunks proven to be stored.
For high $\fileproofNum$, the percentage of distinct file chunks proven grows logarithmically.
For $\fileproofNum = \num{1000}$ (\emph{i.e.}, each proof of possession is computed on $1.6\%$ of the total number of chunks), approximately $150$ timestamps are sufficient to guarantee that $90\%$ of the file is stored correctly.
If the timestamps are $10$ minutes long (\emph{e.g.}, Bitcoin), $\fileproofNum = \num{1000}$ guarantees that the $90\%$ of the file is correctly stored only in $1$ day of protocol execution.
Additionally, our results allow us to determine the type of erasure code to use according to the user's preferences.
For example, a $(0.9\cdot n)$-out-of-$n$ erasure code guarantees the retrievability of the file in $1$ day while a $(n/2)$-out-of-$n$ erasure code reduces the wait time to only $8$ hours.\footnote{For $\fileproofNum = \num{1000}$, $n/2$ chunks are proven in approximately $50$ timestamps (\emph{e.g.,} $8$ hours if each timestamp is $10$ minutes long).}

Based on the results above, we ran a simulation to evaluate the performance of a large instantiation of \systemname\ (large file and several storage-nodes).
The simulation aims to show the impact of the parameter $\ticketproofnum$ (number of proofs accepted at each timestamp).
In more detail, we deployed $1$ Petabyte file composed of approximately $68$ billion chunks (parameter $n$) distributed among $1$ million storage-nodes, each of which entitled to store $\storageblockNum = \num{655360}$ ($10$GB) chunks.
This time, at each timestamp, $\storagenodeNum = \num{400000}$ storage-nodes are challenged on $\num{1024}$ ($16$MB), $\num{8000}$ ($125$MB), $\num{16000}$ ($250$MB), $\num{32000}$ ($500$MB), $\num{48000}$ ($750$MB), and $\num{65536}$ ($1$GB) chunks (parameter $\fileproofNum$).
We ran the simulation with different values of $\ticketproofnum$ (\emph{i.e.}, $\num{1000}$, $\num{10000}$, $\num{50000}$, $\num{100000}$, $\num{200000}$) and we report the number of timestamps needed to prove $90\%$ of the file.


Figure~\ref{fig:test_multiple_l} shows the results.
By increasing $\ticketproofnum$, we can reduce the number of timestamps needed to prove a fixed percentage of the file (in this case, $90\%$).
For example, for $\fileproofNum=\num{8000}$ ($125$MB) and $\ticketproofnum = \num{1000}$, $90\%$ is reached after $\num{19314}$ timestamps.
Instead, for higher values such as $\ticketproofnum = \num{10000}$ and $\ticketproofnum = \num{50000}$, the number of timestamps drops to $\num{2208}$ and $442$, respectively.
Moreover, note that the timestamps and the parameter $\ticketproofnum$ are linearly correlated.
As an example, between $\ticketproofnum = \num{1000}$ and the next order of magnitude $\ticketproofnum = \num{10000}$, the numbers of timestamps ($\num{19314}$ and $\num{2208}$, respectively) differ approximately by the same order.

As already discussed in Section~\ref{par:decentralized-dealer}, $\fileproofNum$ and $\ticketproofnum$ can be tuned to distribute the proof generation overhead among multiple storage-nodes while maintaining (or increasing) the system performance.
Furthermore, the tuning can be adaptively performed by the \systemname\ network (using a similar approach used by Bitcoin to change the PoW difficulty adaptively) according to the network status, \emph{e.g.}, the number of storage-nodes, block creation, and proofs generation time, etc.).

\paragraph*{Success probability of a malicious storage-node}
A malicious storage-node may erase $t$ chunks and still being able to compute a valid proof with a certain probability.
Naturally, this probability depends on the number of stored chunks $\storageblockNum$, the number of deleted chunks $t$, and the number of chunks challenged $\fileproofNum$.
Ateniese et al.~\cite{ateniese2007provable} shows that a malicious storage-node fails to compute a valid proof with a probability $p$ that is:
\[
    1- \left(\frac{\storageblockNum - t}{\storageblockNum}\right)^\fileproofNum \leq p \leq 1- \left( \frac{\storageblockNum - \fileproofNum +1 - t}{\storageblockNum - \fileproofNum +1} \right)^\fileproofNum.
\]
If we set $t$ to be a percentage of $\storageblockNum$, a malicious storage-node fails (with high probability) if it is challenged on a constant number of chunks $\fileproofNum$.
In particular, if $t=1\%$ of $\storageblockNum$, then challenging $\fileproofNum=460$ and $\fileproofNum=300$ chunks permits to achieve $p$ of at least $99\%$ and $95\%$.
We refer the reader to~\cite{ateniese2007provable} for more details.

\paragraph*{Communication complexity}
The communication complexity that \systemname\ adds to the underlying blockchain protocol depends on the PDP scheme used (the output of the $\prove$ algorithm consists in a signature $\sigma$ of size $\secpar$, and a PDP proof $\pi'$).
The publicly verifiable scheme in~\cite{ateniese2007provable} produces proofs of size $O(\log(\fileproofNum) + |\bfile| + v + |N|)$ where $|\bfile|$ is the chunk bit length, $|N|$ is the size of the RSA modulo, and $v$ is the output length of a PRF.
It is reasonable to assume $|\bfile| \gg \log(\fileproofNum)$, $|\bfile| \gg |N|$, $|\bfile| \gg v$, hence the proof size (and thus the additional communication complexity) is mainly determined by $|\bfile|$.
However, the results shown in Figure~\ref{fig:accenture_test} and Figure~\ref{fig:test_multiple_l} depend on the chunk size ($|\bfile|= 16$KB in our experiments).
Therefore, a small $|\bfile|$ would make the communication efficient but decrease the storage guarantees of \systemname.
We observe that it is possible to keep the same guarantees by tuning the parameters $\fileproofNum$ and $|\bfile|$.
To be concrete, if we reduce the chunk size from $16$KB to $1$KB ($16$ times smaller), it is enough to set $\fileproofNum$ to $16\cdot\fileproofNum$ to achieve the same performance while lowering the communication complexity of the protocol by a factor of $16$ (note that the proof size scales logarithmically in $\fileproofNum$).

\section{Conclusions}\label{sec:conclusion}
In this work, we presented \systemname, a blockchain-based decentralized storage system that redefines the current structure of the widely used cloud storage services.
\systemname\ can be built on top of several blockchain systems and uses an augmented network of participants that include storage-nodes and block-creators.

We identified the properties that a decentralized storage system must satisfy, and we provided a detailed comparison between the current state-of-the-art systems and \systemname.
We formally defined \systemname\ and we evaluated its security guarantees.
In addition, we demonstrated the technical feasibility of \systemname\ by implementing a prototype based on the distributed ledger Quorum, and we evaluated its performance.

\bibliographystyle{IEEEtran}
\bibliography{IEEEabrv, biblio}

\ifFULL
\appendices
\section{Proof of theorem~\ref{thm:sec-system}}\label{sec:proof-thm}
Let $t$, $\cStoragenode^*$, $\file = \{\bfile_1, \ldots, \bfile_n\}$ be a timestamp in which the blockchain is extended, the set of $\storagenodeNum$ elected storage-nodes for that timestamp, and an arbitrary file, respectively.
By contradiction, assume that for every set of $\ticketproofnum$ storage-nodes $\{\storagenode^*_i\}_{i \in [\ticketproofnum]} \subseteq \cStoragenode^*$ there exists at least one storage-node $\widetilde{\storagenode_i} \in \{\storagenode^*_i\}_{i \in [\ticketproofnum]}$ that generates a valid proof of possession $\pi$ with a probability non-negligibly close $\delta$ to the probability that the user can extract $\{\bfile_j\}_{j \in \cIndex_{\chal^*_i}}$ by means of a knowledge extractor $\extract$.
Then, we build an attacker $\adversary$ for $\gamePDP_{\pdp, \file, \adversary}$.
$\adversary$ has three random oracles $\hash_1, \hash_2, \hash_3:\bin^*\rightarrow \bin^*$ under its control, and it acts as both the user and dealer for \systemname.
We additionally assume that $\adversary$ can see the messages exchanged between the block-creators and storage-nodes.\footnote{For example, it can join the network without interfering the protocol.}
Through its entire execution, $\adversary$ answers to the queries for the oracles $\hash_1, \hash_2$ in the following way:
\begin{description}
	\item[$\hash_1$:] Upon input $x = (\widehat{\pk}_\storagenode)$, if $(\widehat{\pk}_\storagenode, y) \in \rolist_1$ then return $y$.
	Otherwise, select a random $y' \getsr \bin^*$, add the tuple $(\widehat{\pk}_\storagenode, y')$ to $\rolist_1$, and return $y'$.
	\item[$\hash_2$:] Upon input $x = (\widehat{\selval})$, if $(\widehat{\selval}, y) \in \rolist_2$ then return $y$.
	Otherwise, select a random $y' \getsr \bin^*$, add the tuple $(\widehat{\selval}, y')$ to $\rolist_2$, and return $y'$.
\end{description}
We build $\adversary$ in the following way:
\begin{enumerate}
	\item Sample a random storage-node $\storagenode_j \in \cStoragenode$ and compute the set of chunks indexes $\set{X}_{\storagenode_j}$ that $\storagenode_j$ is entitled to store (as described in $\getFrag$, Construction~\ref{const:costruction}).

	\item Start the game $\gamePDP_{\pdp, \file, \adversary}(\secparam, \fileproofNum, \set{X}_{\storagenode_j})$.

	\item Receive $\pk^*$ from the challenger.

	\item For each $\bfile_i \in \file$, send $\bfile_i$ to the oracle $\tagblock$, and receive the tag $\btag_i$.
	\item Set $\pk^* = \pk_{\subFile}$ and $\encFile = \{\encbFile_i = (\bfile_i, \btag_i)\}_{i \in [n]}$.

	\item Eventually, receive the randomness $r^*$ from the challenger. At this point, $\adversary$ programs the random oracle $\hash_3$ in the following way:
	\begin{description}
		\item[$\hash_3$:]
		Without loss of generality, assume that there are $q=\poly$ queries $x_i = (\pk_{\encFile} || \pk_{\storagenode_j} || \_)$ (\emph{i.e.}, queries with prefix $\pk_{\encFile} || \pk_{\storagenode}$).
		We denote with $\mathcal{Q} = \{x_1, \ldots, x_q\}$ such queries.
		$\adversary$ flips a bit $b \getsr \bin$: If $b=0$, it selects a random index $l \in [q]$, sets $\hash_3(x_{l}) = r^*$, and answers randomly to all other queries.\footnote{$\adversary$ builds a list $\rolist_3$ to answer consistently to the $\hash_3$ queries.}
		Otherwise (i.e, $b=1$), it answers with $\hash_3(x_i) = y_i$ where $y_i\getsr \bin^*$ to each query $x_i$.
	\end{description}

	\item The adversary starts the \systemname\ protocol by publishing $\pk_{\subFile}$ and sending to each storage-node $\storagenode \in \cStoragenode$, with public key $\pk_{\storagenode}$, the file portion $\subFile_\storagenode = \getFrag(\pk_{\encFile}, \encFile, \pk_{\storagenode})$.

	\item Wait until round $t$. Eventually, a leader $\blockcreator^*$ and a set of $\storagenodeNum$ storage-nodes $\cStoragenode^*$ will be elected.
	The leader sends its identification string $\selval$ to each $\storagenode^*_i \in \cStoragenode^*$.
	$\adversary$ finishes to program the random oracle $\hash_3$ in the following way: If $b=1$, it sets $\hash_3(\pk_{\encFile}|| \pk_{\storagenode_j} ||\selval) =r^*$; Otherwise, it sets $\hash_3(\pk_{\encFile}|| \pk_{\storagenode_j} ||\selval)=y$ where $y\getsr\bin^*$.

	\item Eventually, each storage-node $\storagenode^*_i \in \cStoragenode^* $ will output a proof of possession $\pi_i = (\pi'_i, \sigma_i)$.
	$\adversary$ checks the following: if $[(b=0 \land x_{l} \neq (\pk_{\encFile} || \pk_{\storagenode_j} || \selval)) \lor (b=1 \land (\pk_{\encFile} || \pk_{\storagenode_j} || \selval) \in \mathcal{Q})]$; If yes, $\adversary$ aborts.
	Otherwise, it samples a random proof of possession $(\widehat{\pi}', \widehat{\sigma}) = \widehat{\pi} $ generated by a storage-node $\widehat{\storagenode} \in \cStoragenode^*$ and sends $\widehat{\pi}'$ to the challenger.
\end{enumerate}

We start by analyzing the probability of abortion.
Let $E_{\mathsf{abort}}$, $E_1$, $E_2$ be the event that $\adversary$ aborts, $(b=0 \land x_{l} \neq (\pk_{\encFile} || \pk_{\storagenode_j} || \selval))$, and $(b=1 \land (\pk_{\encFile} || \pk_{\storagenode_j} || \selval) \in \mathcal{Q})$, respectively. We can write $\Pr[\lnot E_{\mathsf{abort}}] = 1 - \Pr[E_{\mathsf{abort}}]$ and $\Pr[E_\mathsf{abort}] = \Pr[E_1 \lor E_2] \leq \Pr[E_1] + \Pr[E_2]$.
Assuming that $\Pr[(\pk_{\encFile} || \pk_{\storagenode_j} || \allowbreak\selval) \in \mathcal{Q}] = p_1$ we have:
\begin{align*}
	&\Pr[E_1] = \Pr[b=0] \cdot \Pr[x_{l} \neq (\pk_{\encFile} || \pk_{\storagenode_j} || \selval)]=\\
	&= \frac{1}{2}\bigg(\Pr[x_{l} \neq (\pk_{\encFile} || \pk_{\storagenode_j} || \selval) | (\pk_{\encFile} || \pk_{\storagenode_j} || \selval) \in \mathcal{Q}]\\
	&\qquad \cdot \Pr[(\pk_{\encFile} || \pk_{\storagenode_j} || \selval) \in \mathcal{Q}] \\
	&\qquad + \Pr[x_{l} \neq (\pk_{\encFile} || \pk_{\storagenode_j} || \selval) | \selval \not\in \mathcal{Q}]\\
	&\qquad \cdot \Pr[\selval \not\in \mathcal{Q}]\bigg)\\
	&= \frac{1}{2}[(1-1/q)\cdot p_1 + (1 - p_1)] = \frac{1}{2}(1-p_1/q)
\end{align*}
and $\Pr[E_2] = \Pr[b=1]\cdot\Pr[(\pk_{\encFile} || \pk_{\storagenode_j} || \selval) \in \mathcal{Q}] = \frac{p_1}{2}$.
This allows to conclude that the probability of abortion is:
\begin{align*}
\Pr[E_{\mathsf{abort}}] &\leq \Pr[E_1] + \Pr[E_2] = \frac{1}{2}(1-p_1/q) + \frac{p_1}{2} \\
&= \frac{1}{2}(1-\frac{p_1}{q}+p_1) = \frac{1}{2}(1 + p_1(1-1/q)) \\
&\leq \frac{1}{2}(1 + 1\cdot(1-1/q)) = 1 - \frac{1}{2q}
\end{align*}
Thus, the probability of that $\adversary$ does not abort is $\Pr[\lnot E_{\mathsf{abort}}] \geq \frac{1}{2q}$.
Moreover, by contradiction we know that for every set of of $\ticketproofnum$ storage-nodes $\{\storagenode^*_i\}_{i \in [\ticketproofnum]} \subseteq \cStoragenode^*$ there exists at least one storage-node $\widetilde{\storagenode_i} \in \{\storagenode^*_i\}_{i \in [\ticketproofnum]}$ that generates a valid proof of possession $\pi$ with a probability non-negligibly close $\delta$ to the probability that the user can extract the challenged chunks by means of a knowledge extractor $\extract$.
Hence, $\adversary$ wins the PDP game if and only if $\storagenode_j = \widetilde{\storagenode_i} = \widehat{\storagenode}$ and $\ver_{\pdp}(\pk^*, \chal, \widehat{\pi}') =1$ where $\chal = \genchal_{\pdp}(\fileproofNum, \set{X}_{\storagenode_j}; r^*)$.
We calculate $\Pr[\storagenode_j = \widetilde{\storagenode_i} = \widehat{\storagenode}] = \Pr[\widehat{\storagenode} = \widetilde{\storagenode_i}] \cdot \Pr[\widehat{\storagenode} = \storagenode_j]$ in the following way:
\begin{align*}
&\Pr[\widehat{\storagenode} = \widetilde{\storagenode_i}] \cdot \Pr[\widehat{\storagenode} = \storagenode_j] \geq \frac{1}{\storagenodeNum} \Pr[\widehat{\storagenode} = \storagenode_j] \\
&= \frac{1}{\storagenodeNum}\bigg(\Pr[\widehat{\storagenode} = \storagenode_j | \storagenode_j \in \cStoragenode^*]\cdot \Pr[\storagenode_j \in \cStoragenode^*] \\
&\qquad + \Pr[\widehat{\storagenode} = \storagenode_j | \storagenode_j \notin \cStoragenode^*]\cdot \Pr[\storagenode_j \notin \cStoragenode^*] \bigg)\\
&=\frac{1}{\storagenodeNum}\bigg(\Pr[\widehat{\storagenode} = \storagenode_j | \storagenode_j \in \cStoragenode^*]\cdot (1-\Pr[\storagenode_j \notin \cStoragenode^*]\bigg) \\
&=\frac{1}{\storagenodeNum}\bigg(\frac{1}{|\cStoragenode^*|}\cdot\left(1- \frac{|\cStoragenode|-1}{|\cStoragenode|} \cdot \ldots \cdot \frac{|\cStoragenode|-\storagenodeNum}{|\cStoragenode| - \storagenodeNum + 1}\right) \bigg)\\
&= \frac{1}{\storagenodeNum^2}\bigg(1- \frac{|\cStoragenode|-\storagenodeNum}{|\cStoragenode|}\bigg) = \frac{1}{\storagenodeNum^2}\cdot\frac{\storagenodeNum}{|\cStoragenode|} = \frac{1}{\storagenodeNum\cdot|\cStoragenode|},
\end{align*}
where we used the fact that the set of $\storagenodeNum$ elected storage-nodes $\cStoragenode^*$ is randomly computed by hashing the unpredictable identification string $\selval$.

Conditioned on $\lnot E_{\mathsf{abort}} \land (\storagenode_j = \widetilde{\storagenode_i} = \widehat{\storagenode})$ and since at timestamp $t$ the blockchain is extended, we have $\ver_{\pdp}(\pk^*, \chal^*_j, \allowbreak\pi'_j)= 1$ where the challenge $\chal^*_j$ is computed in the following way:
\begin{align*}
	\chal^*_j &= \genchal_{\pdp}(\fileproofNum,;\hash_3(\pk_{\encFile} || \pk_{\storagenode_j} || \selval)) \\
	&= \genchal_{\pdp}(\fileproofNum,;\hash_3(\pk^* || \pk_{\storagenode_j} || \selval))\\
	&= \genchal_{\pdp}(\fileproofNum, \set{X}_{\storagenode_j};r^*).
\end{align*}
Hence, $\pi'_j = \widehat{\pi'}$ is a valid proof of possession for the game $\gamePDP_{\pdp, \file, \adversary}(\allowbreak\secparam, \fileproofNum, \set{X}_{\storagenode_j})$ with probability greater than $\delta \cdot \frac{1}{2q} \cdot \frac{1}{\storagenodeNum\cdot|\cStoragenode|}$ where $|\cStoragenode|$ and $\storagenodeNum$ are positive constants.\footnote{The total number of storage-nodes $|\cStoragenode|$ and $\storagenodeNum$ are independent from the security parameter $\secpar$.}
This concludes the proof.

\fi

\end{document}